\documentclass[aps,pra,twocolumn,superscriptaddress,english,showpacs]{revtex4-1}

\usepackage{lipsum}
\usepackage{natbib}
\usepackage[utf8]{inputenc}
\usepackage[T1]{fontenc}
\usepackage[english, french]{babel} 
\usepackage{amsmath}
\usepackage{amsfonts}
\usepackage{makeidx}
\usepackage{graphicx}
\usepackage{mathtools} 
\usepackage{braket} 
\usepackage[colorlinks=true, linkcolor=black, citecolor=black]{hyperref} 
\usepackage[usenames,dvipsnames]{xcolor} 

\usepackage{tikz} 
\usetikzlibrary{positioning}
\usepackage{ dsfont } 
\usepackage{physics}
\usepackage{xifthen}

\usepackage{subfiles}

\usepackage{standalone}

\usepackage{subcaption}

\usepackage{psfrag}

\renewcommand{\d}[1]{\mathrm{d}#1}

\newcommand{\om}{\ensuremath{\omega}}
\DeclareMathOperator{\loc}{C}
\DeclareMathOperator{\sub}{S}
\newcommand{\idos}{\ensuremath{\text{idos}}}
\newcommand{\hc}{\ensuremath{\text{H.c.}}}
\newcommand{\reg}{\ensuremath{\mathcal{R}}}
\newcommand{\gen}{\ensuremath{Z}}

\newcommand{\AB}{Ammann-Beenker}
\newcommand{\SKK}{SKK}
\newcommand{\simop}[1]{\ensuremath{\underset{#1}{\sim}}}
\newcommand{\supp}{\ensuremath{\Delta_f}}
\newcommand{\psivar}{\ensuremath{\psi_\text{var}}}
\newcommand{\locvar}{\ensuremath{C_\text{var}}}

\newcommand{\nperp}[2]%
{%
		\ifthenelse{\isempty{#1}}{\ensuremath{\widetilde{x}_\perp}}{\ensuremath{\widetilde{x}_\perp(#1,#2)}}%
}

\newcommand{\dlogpsi}{\ensuremath{\omega}}

\begin{document}


 


\title{Critical eigenstates and their properties in one and two dimensional quasicrystals.}
\author{Nicolas Macé $^1$, Anuradha Jagannathan$^1$, Pavel Kalugin$^1$, Rémy Mosseri$^2$ and Frédéric Piéchon$^1$\\
$^1${Laboratoire de Physique des Solides, Université Paris-Saclay, 91400 Orsay, France}\\
$^2${Laboratoire de Physique Théorique de la Matière Condensée, Universit\'e Pierre et Marie Curie, 75005 Paris, France}
}
\selectlanguage{english}

\date{\today}

\begin{abstract} 
We present exact solutions for some eigenstates of hopping models on one and two dimensional quasiperiodic tilings and show that they are ``critical" states, by explicitly computing their multifractal spectra. 
These eigenstates are shown to be generically present in 1D quasiperiodic chains, of which the Fibonacci chain is a special case. We then describe properties of the ground states for a class of tight-binding Hamiltonians on the 2D Penrose and \AB\ tilings.
Exact and numerical solutions are seen to be in good agreement. 
\end{abstract}

\maketitle

\begin{figure}[htp]
\centering
\includegraphics[width=0.5\textwidth]{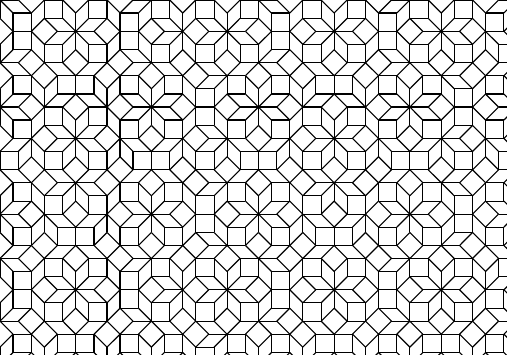}
\caption{A patch of the \AB\ tiling.}
\label{fig:AB_patch}
\end{figure}

In quasicrystals, the arrangement of atoms is non-periodic yet long-range ordered. This type of quasiperiodic order is best illustrated by considering tilings. These are structures obtained by packing a small number of basic elements, or tiles, according to certain specified rules.
Quasiperiodic tilings can exhibit non-crystallographic symmetries: in this article we will focus on the \AB\ tiling (fig.~\eqref{fig:AB_patch}) that has a eight-fold orientational symmetry, and on the celebrated Penrose tiling which has a five-fold symmetry.

Single-electron properties of 1D quasicrystals are rather well understood: many spectral properties of quasiperiodic chains are known, and the eigenstates are also well characterized.
In contrast, even the simplest models of 2 and 3 dimensional quasicrystals have resisted theoretical investigations.
Consider for example the \AB\ tiling (fig.~\eqref{fig:AB_patch}), and a  tight-binding model for non-interacting electrons hopping between nearest neighbor vertices on this tiling:
\begin{equation}
\label{eq:SimpleHam}
	H = -t\sum_{\langle i,j \rangle} c_j^\dagger c_i + \hc
\end{equation}
Although this Hamiltonian is simple, no solutions were known for any of its eigenstates, apart from trivial confined eigenstates at the middle of the spectrum \cite{LocalizedKohmoto}.

This situation changed recently, when Kalugin and Katz \cite{KaluginKatz}, building on the work of Sutherland \cite{Sutherland}, deduced the form of the ground state of the Hamiltonian \eqref{eq:SimpleHam} on the Penrose and Ammann-Beenker tilings. 

Taking their cue from periodic crystals, where Bloch states are given by $\psi_k(r) = u_k(r) e^{ik.r}$ with $u$ a periodic function, Kalugin and Katz \cite{KaluginKatz} proposed that the ground state of the model \eqref{eq:SimpleHam} be given by a product of two factors, namely
\begin{equation}
\label{eq:SKK}
	\psi(i) = \loc(i) e^{\kappa h(i)}
\end{equation}
In this expression, $\kappa$ is a real constant (note that in \cite{KaluginKatz} the notation $\lambda = \exp(2\kappa)$ is used).
The pre-exponential factor $\loc(i)$ of the ansatz \eqref{eq:SKK} is a quasiperiodic function, a site-dependent amplitude which depends only on the arrangement of the atoms around the site $i$. In other words, $\loc(i) \simeq \loc(j)$ if the arrangement of the atoms around site $i$ matches the arrangement of the atoms around site $j$ out to a large distance. 
The non-local \emph{height field} $h(i)$ in the exponential depends on the geometrical properties of the tiling. We will refer to eigenstates of this form henceforth as Sutherland-Kalugin-Katz -- or \SKK\ eigenstates.

In this article, we consider generalizations of the results of Kalugin and Katz to other tight-binding models. We consider first the relatively simple case of models on 1D quasiperiodic chains, and show that they admit eigenstates of a form similar to that given in Eq.~\eqref{eq:SKK}. We next consider the 2D case, for the \AB\ and Penrose tilings. and we show that the ground states continue to have the \SKK\ structure even when the Hamiltonian in Eq.\eqref{eq:SimpleHam} is generalized to include onsite potentials.
We show that these eigenstates are \emph{critical}, i.e. neither localized nor extended, by an explicit analytical calculation of generalized fractal dimensions. These exact results provide confirmation of a long-held surmise based on numerical calculations \cite{DimsSchreiber}. 

This article is organized as follows.
In part I, we present exact solutions for the one-dimensional case and their associated properties.
In part II, we consider models on two dimensional tilings, namely the Ammann-Beenker and Penrose tilings. We show that they are \SKK\  eigenstates, whose multifractal properties are calculated and compared with numerical data.


\section{\SKK\ states on 1D chains and their properties}

In this section, we will focus on one-dimensional models, which provide a good starting point in the study of \SKK-like eigenstates. Hopping models  on a number of different aperiodic chains will be discussed.
For definiteness, we begin with the example of the Fibonacci chain.

\subsection{The Fibonacci chain and hopping Hamiltonian}


It will be useful in the following discussion to introduce some notation, along with a reminder of basic definitions. Let us introduce the Fibonacci substitution rule $\sub$, acting on the two letters alphabet $\mathbf{\mathcal{A}} = \{a, b\}$:
\begin{equation}
\label{eq:FiboSub}
	\sub: 
	\begin{cases}
		a \to ab\\
		b \to a.
	\end{cases}
\end{equation}
Letting the substitution act repeatedly on the letter $b$ generates a sequence of words  $C_l = \sub^l(b)$ of increasing length.
Note that the lengths of the words are Fibonacci numbers.
In the limit $ l \to \infty$, one obtains an infinite Fibonacci word (see \cite{Baake2013} for more details on symbolic substitutions).

To the Fibonacci word we associate a collection of hopping amplitudes $\{t_i\}$, in the following way:
\begin{equation}
	t_i =  
	\begin{cases}
		t_a & \text{if the letter $i$ is $a$}\\
		t_b & \text{if the letter $i$ is $b$}.
	\end{cases}
\end{equation}
The Fibonacci Hamiltonian is then built using this sequence of hoppings:
\begin{equation}
\label{eq:FiboHam}
	H = \sum_{i} t_i c_{i+1}^\dagger c_i + \hc
\end{equation}

\begin{figure}[htp]
\centering
\begin{subfigure}[b]{0.4\textwidth}
	\includegraphics[width=\textwidth]{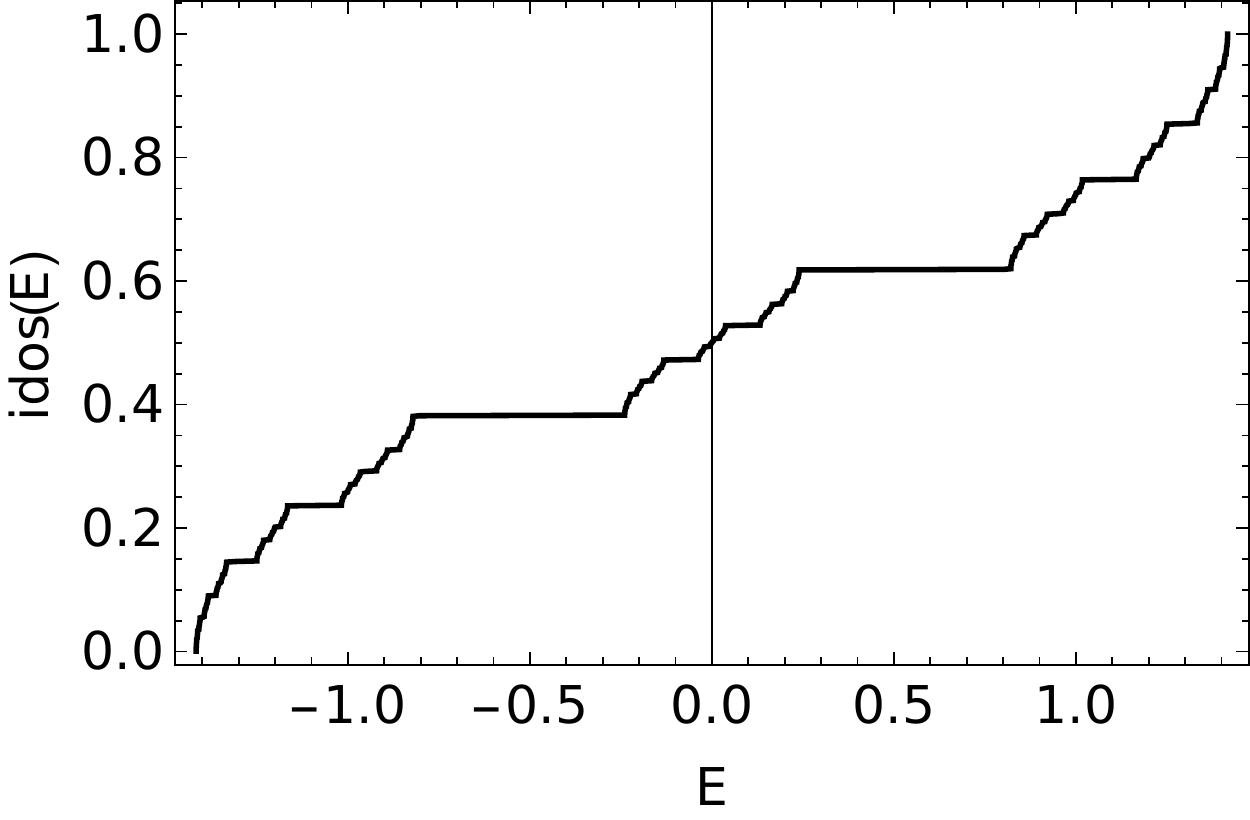}
	\caption{The integrated density of states of the model \eqref{eq:FiboHam} with $\rho = 0.5$. }
\end{subfigure}
\begin{subfigure}[b]{0.4\textwidth}
	\includegraphics[width=\textwidth]{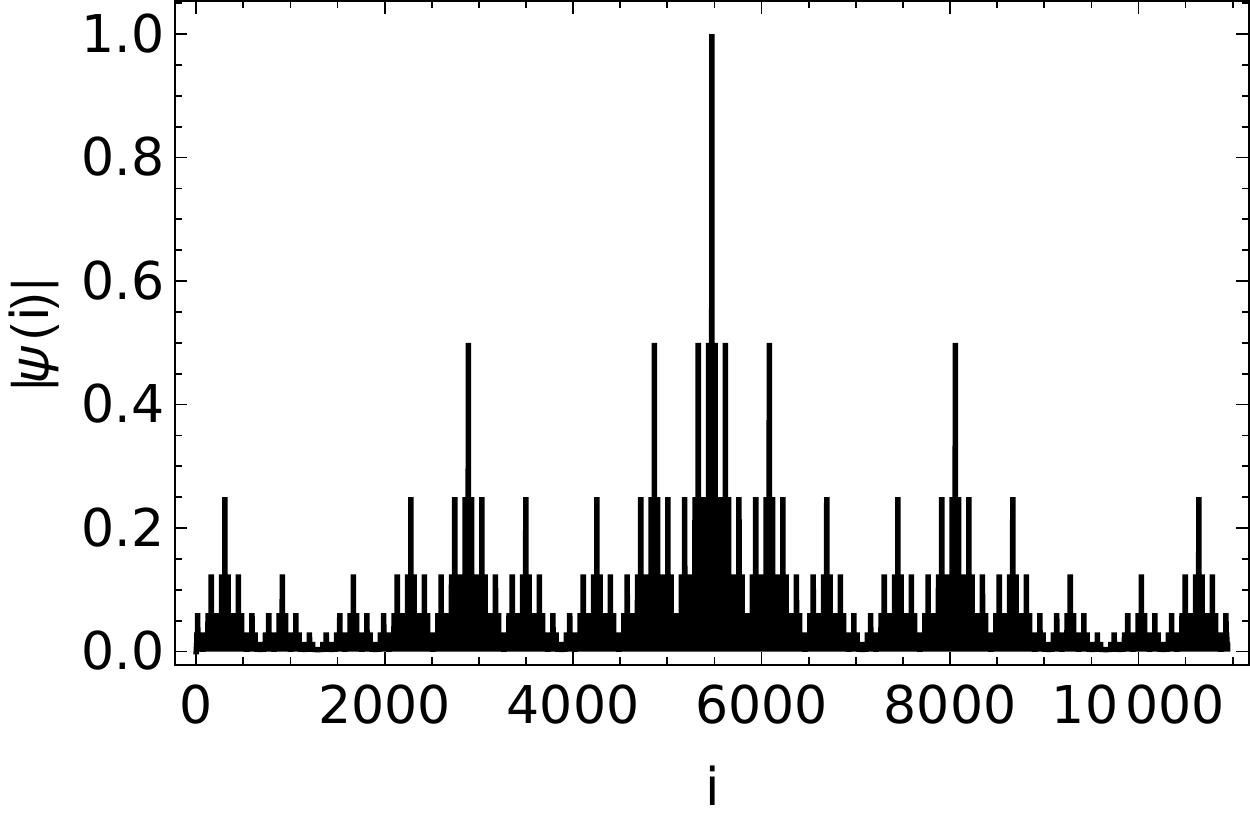}
	\caption{Absolute value of wavefunction amplitudes in the $E=0$ state.}
\end{subfigure}
\caption{}
\label{fig:central_state_fibo}
\end{figure}

In anticipation of the two-dimensional model of Part II,  and at the risk of belaboring the obvious, we note that the letters $a$ and $b$ play the role of 1D tiles, and that ``atoms'' are located in between the tiles.
Up to a global rescaling of the energies, the only free parameter of this Hamiltonian is the ratio of the two hopping amplitudes $\rho = t_A/t_B$.
If $\rho = 1$ the chain is periodic. If $\rho = 0$ or $\rho = \infty$ the chain is a collection of decoupled molecules. 
We will exclude these 3 trivial cases and assume $\rho \neq 0,1,\infty$ henceforth.

It is well known that the spectrum of the Fibonacci Hamiltonian is fractal \cite{SutoCantorSet}, as illustrated in the top panel of fig.~\eqref{fig:central_state_fibo} which shows the integrated density of states, $\idos$(E) defined by the fraction of states of energy smaller than $E$. The fractal dimensions of the spectrum can be computed in the limits $\rho \sim 1$ \cite{RudingerSireSpec} and $\rho \ll 1$ \cite{PiechonSpec}. The structure of the eigenstates is also well understood in these two limits \cite{SireStates, NiuNori, ThiemStates, Mace}.
Away from these limits, however, the structure of the eigenstates is not known, with the notable exception of the $E=0$ state at the center of the spectrum shown in fig.~\eqref{fig:central_state_fibo}. This state is of the \SKK\ type, as we will discuss in the next section.

\subsubsection{The $E=0$ eigenstate}
The Hamiltonian \eqref{eq:FiboHam} is particle-hole symmetric, i.e., to an eigenstate at energy $E$ corresponds an eigenstate at energy $-E$, which is related to the former by a sign change on the subchain of even (or odd) sites.
In particular, the central state at $\text{idos} = 1/2$, which we wish to study, has energy $E=0$ and is doubly degenerate. 
We recall that this state can be built using the so-called trace-map method, used for instance in \cite{FujiwaraExact}  to obtain a description of this state, and in particular to compute its fractal dimensions.
The fractal dimensions of this state can also be computed exactly using a renormalization-group approach \cite{Mace}. We wish to show now that the $E=0$ state is an \SKK\ like state \eqref{eq:SKK}.
To do this, we first introduce a decoration of the chain by \emph{arrows} and a \emph{height field} which is the integral of the arrows.

\subsubsection{Arrows and height field}
The tight-binding equations \eqref{eq:FiboHam} for the central state are
\begin{equation}
\label{eq:tb_fibo}
	t_{i+1} \psi_{i+2} + t_{i} \psi_{i} = 0
\end{equation}
Even and odd subchains are seen to be completely decoupled, so that one of the two $E=0$ eigenstates can be chosen to vanish on the odd subchain, and the other state to vanish on the even subchain. We focus on the first of these eigenstates, designated by $\psi$.  
By symmetry all the same properties will also hold for the state living on the odd sites.

Rewriting the tight-binding equations \eqref{eq:tb_fibo},  we have
\begin{equation}
\label{eq:hop2sites}
	\psi(2(i+1)) = -\rho^{A(2i \to 2(i+1))} \psi(2i)
\end{equation}
where we have introduced the \emph{arrow function} $A$, defined on \emph{pairs} of bonds, as follows
\begin{equation}
\label{eq:arrow}
	A:
	\begin{cases}
		A(ab) & = +1 \\
		A(ba) & = -1 \\
		A(aa) & = 0 \\
		A(bb) & = 0
	\end{cases}
\end{equation}
where the last case of two consecutive $b$-bonds, not present in the Fibonacci chain, is included for future generalizations.
\begin{figure}
	\centering
   \begin{subfigure}[b]{0.4\textwidth}   		\includegraphics[width=\textwidth]{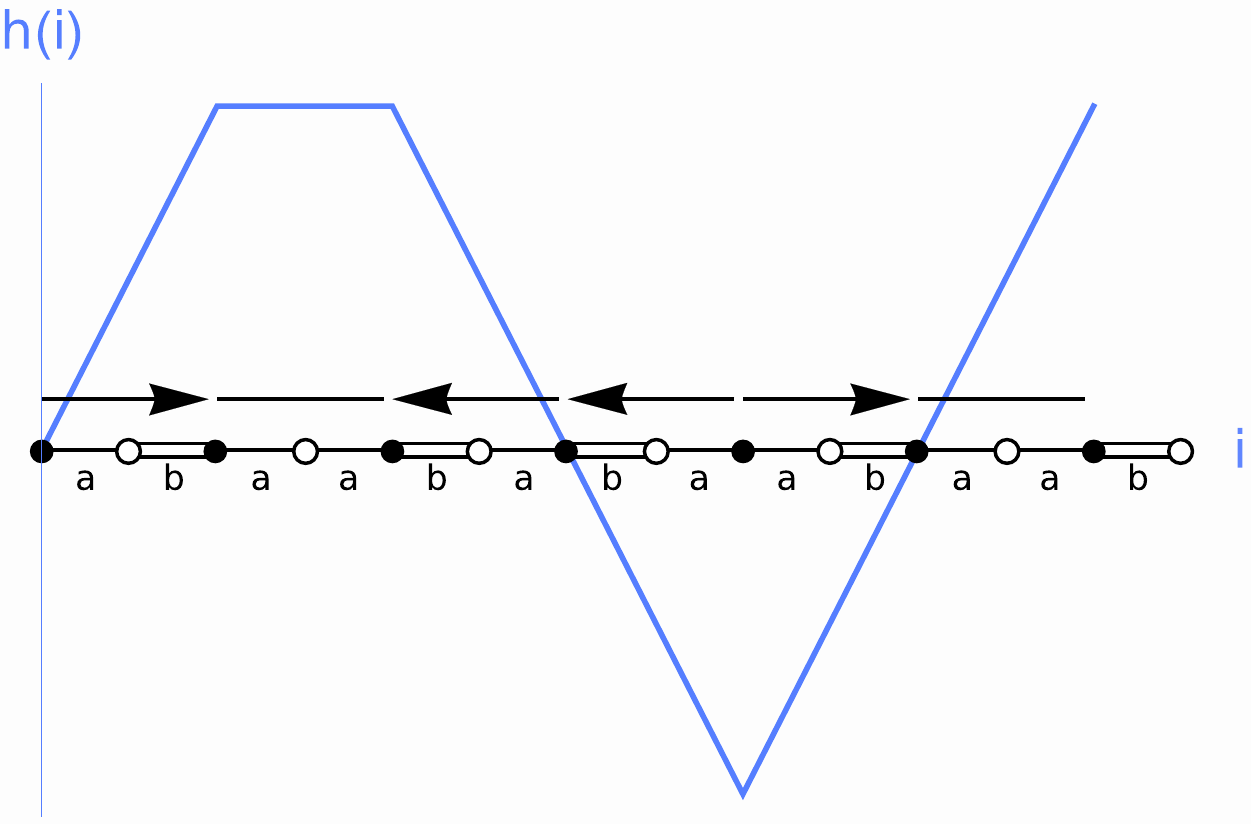}
   		\caption{A segment of the Fibonacci chain showing a sequence of $t_A$ (single bond) and $t_B$ (double bond) couplings. The arrow field and the associated height field are shown for one of the subchains (the blue line is drawn to guide the eye).}
   \end{subfigure}
   \vskip 0.5cm
   \begin{subfigure}[b]{0.4\textwidth}
   	    \includegraphics[width=\textwidth]{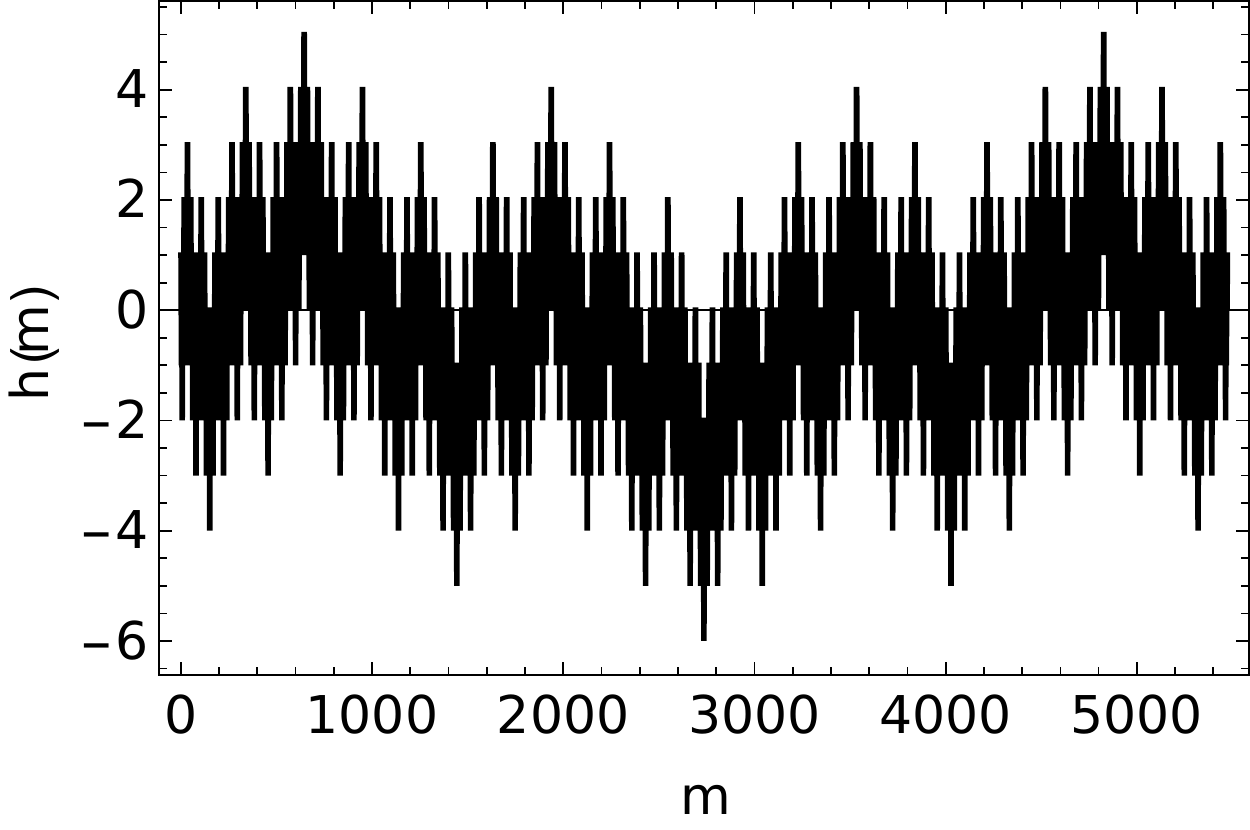}
        \caption{The height field on a larger piece of the Fibonacci chain.}
   \end{subfigure}
	\caption{}
	\label{fig:arrows_heights}
\end{figure}
The name ``arrow'' refers to the fact that one can associate to each pair of bonds an arrow pointing to the right (+1), or to the left (-1) or neither (0), cf fig.~\eqref{fig:arrows_heights}.

Iterating the relation \eqref{eq:hop2sites}, one obtains the wave-function in terms of a height function, $h(m)$
\begin{equation}
\label{eq:FiboState}
	\psi(2 m) = (-1)^m \rho^{h(m)} \psi(0)
\end{equation}
where 
\begin{equation}
	h(m) = \sum_{0 \leq i < m} A(2i \rightarrow 2(i+1))
\end{equation}
The height function is thus the integral of the arrow field (see fig.~\eqref{fig:arrows_heights}) which is itself determined by the quasiperiodic structure of the chain. 
Comparing eq.~\eqref{eq:FiboState} with eq.~\eqref{eq:SKK}, one sees that
the central state itself has not the \SKK\ form, but writes as the product of an \SKK\ and a periodic function:
\begin{equation}
\label{eq:SKKoneD}
	\psi(2m) = (-1)^m \exp(\kappa h(2m))
\end{equation}
with $\kappa = \ln(\rho)$. In the periodic limit $\rho \to 1$, one recovers the Bloch state $\psi(2m) = (-1)^m$.

Note that the reasoning leading to Eq.~\eqref{eq:FiboState} is valid for an {\em arbitrary} sequence of couplings, random or nonrandom.
It is instructive to consider an example of a disordered chain for which the distribution of $a$-bonds and $b$-bonds is random. 
Consider to fix the ideas the case of $a$ and $b$ bonds distributed randomly, independently and in equal proportion along the chain.
Then the arrow function is randomly distributed, with zero mean $\langle A \rangle = 0$, and the distribution of heights for the random chain is Gaussian, by virtue of the central limit theorem.
The typical height corresponds to the standard deviation which is 
\begin{equation}
	\sigma^{rand}(m) \simop{m \to \infty} \sqrt{m}  
\end{equation}
to leading order in $m$.
Eq.~\eqref{eq:FiboState} then implies that the wavefunction must have a stretched exponential form, since typical amplitude on the $2m^\text{th}$ site  $|\psi(2m)| \sim e^{-cst\sqrt{m}}$. This form, first shown for off-diagonal disordered models in \cite{Offdiagdisorder}, is particular to the zero energy state, all other states being exponentially localized.
The $E=0$ state of the random system is very different from that of the quasiperiodic chain, where the typical height on site $m$ scales as $\sqrt{\log m}$, as we will show next.

\subsubsection{Diffusion equation for the height function}
\label{sec:1Dheight}
In the following, we consider how to characterize the height field for the hopping model on the Fibonacci chain.
Let us define $r = ab$ (letter corresponding to a right arrow), $l = ba$ (letter corresponding to a left arrow), and $u = aa$ (letter corresponding to no arrow).
In the case of the Fibonacci chain, as we noted before, the group $bb$ never occurs.

The Fibonacci substitution rule \eqref{eq:FiboSub} iterated three times results in an effective substitution rule $\sub'$ for the arrows:
\begin{equation}
\label{eq:ArrowSub}
	\sub': 
	\begin{cases}
		r \to rull\\
		l \to rulr \\
		u \to rullr
	\end{cases}
\end{equation}
Let $\mathbf{v} = (N_r, N_l, N_u)$ be a vector whose entries are respectively the number of right, left and null arrows on a given chain. Upon inflation of this chain \eqref{eq:ArrowSub},  it is transformed to $M \mathbf{v}$  where 
\begin{equation}
\label{eq:inflationmat}
	M = 
	\begin{bmatrix}
		1 & 2 & 2 \\
		2 & 1 & 2 \\
		1 & 1 & 1
	\end{bmatrix}.
\end{equation}The \emph{inflation matrix} $M$ provides information about the distribution of arrows. 
In particular, the relative frequencies of the arrows are given by the Perron–Frobenius eigenvector (i.e. the eigenvector associated with the largest eigenvalue).
Here, one finds
\begin{equation}
\label{eq:freqs}
		\begin{cases}
		f(r) &= \tau^{-2}\\
		f(l) &= \tau^{-2} \\
		f(u) &= \tau^{-3},
	\end{cases}
\end{equation}
where $\tau$ is the golden ratio.

The substitution $\sub'$ replaces a given arrow by a block of arrows.
Notice that the effective arrow corresponding to this superblock is reversed compared to the original one.
In other words, the substitution $\sub'$ preserves the heights of the initial and final sites {\it up an overall sign}. Thus, under a single application of $\sub'$, sites of large (resp. small) wavefunction amplitude switch roles. Wavefunction amplitudes are left invariant when the substitution is applied an even number of times.
This observation leads one to conclude that the wavefunction is probably delocalized, since
the substitution acts to ``push sites further apart''. 
Thus, the eigenstate has non-zero amplitude on sites that can be pushed arbitrarily far apart, and hence is not a localized state.

Consider now the sequence of regions of larger and larger size obtained by applying the substitution $\sub'$ to an original region $\reg_0$:
\begin{equation}
\label{eq:seq}
	\reg_0 \xrightarrow[\sub']{} \reg_1 \xrightarrow[\sub']{} \dots \xrightarrow[\sub']{} \reg_{t} \xrightarrow[\sub']{} \reg_{t+1} \xrightarrow[\sub']{} \dots
\end{equation}
Let us now focus on the height distribution function, which gives the number of times the height $h$ is reached inside the region $\reg_t$, and is relevant for the computation of the fractal dimensions of the central ($E=0$) state.
One can define environment-specific distributions, $N_\mu^{(t)}(h)$, where $\mu$ denotes the local environment. There are three different local environments on the Fibonacci chain which we define according to the nature of the arrow immediately following the site, namely, $\mu = r, ~l, ~u$, appearing with the frequencies \eqref{eq:freqs}. The  
$N_\mu^{(t)}(h)$ give the number of times height $h$ is found on the local environment $\mu$ in region $\reg_t$. 
They are the three components of the vector $\mathbf{N}^{(t)}(h)$ which evolves with each inflation step.

We now write the recurrence formula which relates the vector $\mathbf{N}^{(t)}$ to $\mathbf {N}^{(t+1)}$. Recall that, after one inflation, the new heights are equal to the previous heights plus a shift of $\pm 1$ or 0, and they undergo a change of sign. Introducing a \emph{generalized inflation matrix} that operates on the heights, as discussed in \cite{Sutherland2, Tokihiro, Repetowicz}, we find 
\begin{equation}
\label{eq:FP}
	\mathbf{N}^{(t+1)}(-h) = \sum_{h'= -1}^1 M(h') \mathbf{N}^{(t)}(h-h'),
\end{equation}
where the $3\times 3$ $M(h')$ matrices are termed generalized inflation matrices. $M_{\mu,\nu}(h)$ counts the number of times the environment $\mu$ corresponding to height $h$ appears in $\sub'(\nu)$, the inflation of the environment $\nu$ associated to height 0. Their explicit expressions are given by
\begin{align}
	M(-1) &= 	\begin{bmatrix}
		0 & 0 & 1 \\
		0 & 0 & 0 \\
		0 & 0 & 0
	\end{bmatrix} \\
    M(0) &= 	\begin{bmatrix}
		1 & 2 & 1 \\
		1 & 0 & 1 \\
		0 & 0 & 0
	\end{bmatrix} \\
    M(1) &=	\begin{bmatrix}
		0 & 0 & 0 \\
		1 & 1 & 1 \\
		1 & 1 & 1
	\end{bmatrix}
\end{align}
Note that the sum of the generalized inflation matrices, $M = \sum M(h')$, is just the inflation matrix $M$ of Eq.~\eqref{eq:inflationmat}.

Eq.~\eqref{eq:FP} is a Fokker-Planck-like equation, in which the number of inflations, $t$, plays the role of time, the height $h$ plays the role of a spatial variable, and the generalized inflation matrices $M(\delta h)$ are transition rates.
As such, one expects that the height frequency $P_\mu^{(t)}(h) = N_\mu^{(t)}(h)/\sum_h N_\mu^{(t)}(h)$  will converge to a Gaussian form in the large time limit. 
We prove that it is indeed the case for the Fibonacci chain, as well as for general substitutions under reasonable assumptions.

In order to do this, we introduce the generating function of the probability distribution (the \emph{partition function}):
\begin{equation}
\label{eq:partition}
	\gen^{(t)}_\mu(\beta) = \sum_{h \in \mathbb{Z}} N_\mu^{(t)}(h)e^{\beta h}
\end{equation}
The evolution equation \eqref{eq:FP} is recast to 
\begin{equation}
	\mathbf{\gen}^{(t+2)}(\beta) = \tilde{M}(-\beta)\tilde{M}(\beta) \mathbf{\gen}^{(t)}(\beta)
\end{equation}
where $\tilde{M}(\beta) = \sum_{h} M(h) \exp(-\beta h)$, and where we iterated twice to take care of the sign change.
From this recursion relation, and since all the coefficients of $\tilde{M}$ are strictly positive, the Perron–Frobenius theorem applies, and the large time behavior of the partition function must be of the form
\begin{equation}
\label{eq:PartitionFunctionScaling}
	\gen_\mu^{(2t)}(\beta) \simop{t \to \infty} \omega^t(\beta) f_\mu(\beta)
\end{equation}
where $\om(\beta)$ is the largest eigenvalue of $\tilde{M}(-\beta) \tilde{M}(\beta)$, and $f_\mu(\beta)$ is the associated eigenvector.  
Explicit calculation gives
\begin{equation}
\label{eq:OmFibo}
	\om(\beta) = \left( \frac{(1 + e^\beta)^2 + \sqrt{(1 + e^\beta)^4 + 4 e^{2\beta}}}{2 e^\beta} \right)^2.
\end{equation}
Thus, in the $t \to \infty$ limit, the distributions all converge to the Gaussian distribution:
\begin{equation}
	P^{(t)}_\mu(h) \sim \frac{f_\mu}{\sqrt{4 \pi D t}} \exp\left(-\frac{h^2}{4 D t}\right)
\end{equation}
where $D$ -- which we call the diffusion coefficient by analogy with a diffusion process -- is given by
\begin{equation}
	D = \frac{1}{4} \frac{\partial^2 \log \omega}{\partial \beta^2} \Bigg|_{\beta=0} = \frac{1}{2\sqrt{5}}
\end{equation}
The typical height is given by the standard deviation, and grows like $\sigma^\text{Fib}(y) \sim \sqrt{2Dt}$, while the length of the chain, after $t$ inflations is $L^{(t)} \sim \tau^{3t} L^{(0)}$. 
Thus, on a piece of tiling of size $L$, the typical height scales as $\sigma^\text{Fib}(L) \sim \sqrt{\log L}$. The wavefunction amplitudes should therefore typically decrease more slowly than for the random case we discussed earlier.
We will show in the next section that the eigenstate is in fact a critical state, by explicitly computing its fractal dimensions.

A few comments can be made at this point: we note that the scaling of the partition function \eqref{eq:PartitionFunctionScaling} and the Gaussian nature of the height distribution are generic properties. They will hold for any height field on any substitution tiling, provided that the height is invariant (up to a sign) under inflation, and provided that the entries of the corresponding inflation matrix $\tilde{M}$ are strictly positive.
These two properties hold for the 1D and 2D quasiperiodic substitution tilings we will discuss below, namely, the series of metallic mean chains in 1D, and the Penrose and Ammann-Beenker tilings in 2D. Only the particular form of $\omega(\beta)$ varies, depending on the tiling and the height field. In contrast, these properties do not hold for the b3 chain which is not quasiperiodic (see below).

Additionally, note that although the arrows can be constructed from the local environments, their integral, the height field, is distributed in an environment-independent fashion
\footnote {This non-trivial fact can be understood in the context of the tiling cohomology :the height function is an element of the first cohomology group of the tiling, and as such, it cannot depend on the local environment \cite{Sadun2014}.}.
This property implies that the fractal dimensions of the \SKK\ eigenstates \eqref{eq:SKK} depend only on the distribution of heights, as we will now see.

\subsubsection{Fractal dimensions of the $E=0$ wave function}
Fractal dimensions are a way to completely and compactly characterize fractal sets such as the eigenstates of quasiperiodic tilings \cite{GeneralizedDims, FractalsHalsey}. 
The fractal dimensions of a wavefunction determine whether the state is localized, extended or critical (as discussed for example in \cite{KohmotoCriticalWF, FujiwaraExact, HiramotoExact, DimsSchreiber}).
In the context of Anderson models, the fractal dimensions can be related to the scaling exponents of the critical point, see e.g.~\cite{MirlinAndersonExact}.
The fractal dimensions can be related, in certain cases, to exponents describing the wavepacket dynamics \cite{PiechonDiffusion}.

We recall for completeness the definition of the fractal dimensions in the context of the wavefunctions (see \cite{FractalsHalsey, Janssen}) for more detailed discussions).
We first introduce the $q$-weight of the wavefunction $\psi$:
\begin{equation}
\label{eq:chi}
	\chi_q(\psi, \reg) = \frac{\sum_{i \in \reg} |\psi(i)|^{2q}}{\left(\sum_{i \in \reg} |\psi(i)|^2\right)^q}
\end{equation}
where the sums run over all sites in a given region $\reg$.
The $q$-weight is a measure of the fraction of the presence probability contained inside region $\reg$. 

As previously, we consider a sequence of regions $\reg_t$ whose size grows to infinity.
The $q^\text{th}$ fractal dimension, $d_q(\psi)$, is the scaling of the $q$-weight with the volume of the region:
\begin{equation}
\label{eq:dq}
	d_q(\psi) = \lim_{t \to \infty} \frac{-1}{q-1} \frac{\log \chi_q(\psi, \reg_t)}{\log \Omega}
\end{equation}
where $\Omega$ is the volume (number of sites) inside region $\reg$.

The Legendre transform of the fractal dimensions is the so-called multifractal spectrum.
More precisely, one defines
\begin{equation}
	\alpha_q = \frac{\d \tau_q}{\d q}
\end{equation}
and
\begin{equation}
\label{eq:legendre}
	f(\alpha_q) = q \alpha_q - \tau_q,
\end{equation}
where $\tau_q = (q-1)d_q$.
The $f(\alpha)$ spectrum has a straightforward physical meaning: $f(\alpha)$ gives the fraction of sites around which the wavefunction has scaling $\alpha$ . 
In our case, a trivial (i.e. reduced to a point) multifractal spectrum is the signature of an extended or localized eigenstate \footnote{One can indeed show that for a localized state, the multifractal spectrum is the point $(\alpha, f(\alpha)) = (0, 0)$ while for an extended state, the spectrum is the point $(\alpha, f(\alpha)) = (1, 1)$.}, while a non-trivial multifractal spectrum is the signature of a critical eigenstate.

The computation of the fractal dimensions can be carried out for a state of the \SKK\ form. We find (see the appendix for the details)
\begin{equation}
	d_q(\psi) = \frac{1}{q-1} \log\left( \frac{\om(2 \kappa)^q}{\om(2 \kappa q)} \right) / \log(\omega(0)).
\end{equation}

The $f(\alpha)$ spectrum can be computed exactly for the $E=0$ states. 
Letting $x = 2\kappa$, we have
\begin{equation}
	\alpha_q = \log(\omega(x)) - x \frac{\omega'(qx)}{\omega(qx)}
\end{equation}
and
\begin{equation}
	f(\alpha_q) = \log(\omega(qx)) - qx \frac{\omega'(qx)}{\omega(qx)}
\end{equation}
\begin{figure}
	\centering
	\includegraphics[scale=0.9]{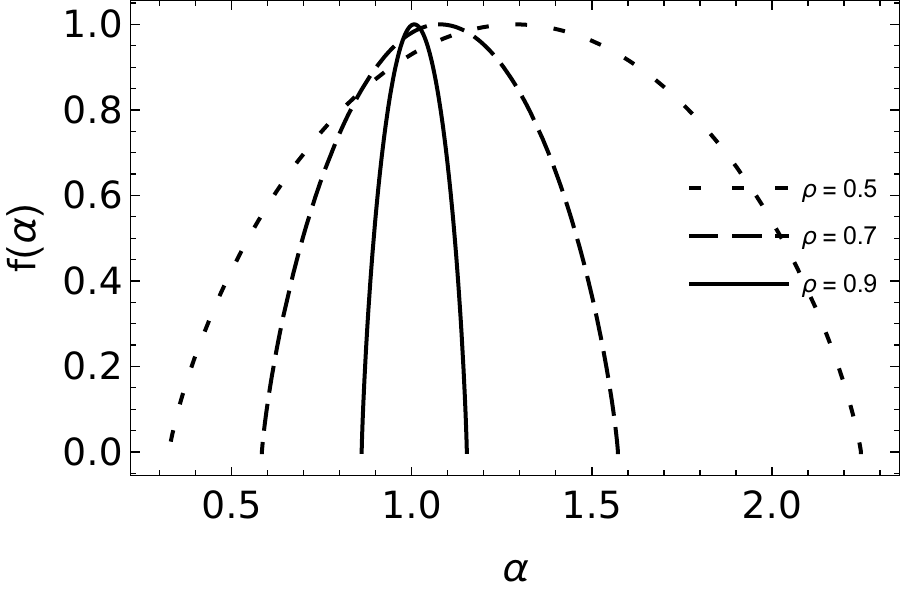}
	\caption{The multifractal spectrum of the central state of the Fibonacci chain, for different value of the coupling ratio $\rho$.  }
	\label{fig:multifractal_spectrum_fibo}
\end{figure}

Figure \eqref{fig:multifractal_spectrum_fibo} shows the multifractal spectrum of the central state of the Fibonacci chain, for different values of the ratio of the couplings, $\rho$.
The spectrum is clearly multifractal (not reduced to a single point), meaning that the $E=0$ state is critical, as we expected.
Note that as $\rho \to 1$, the support of the spectrum becomes narrower, meaning that the state approaches the limiting Bloch wave form in the periodic limit. 
We also remark that the multifractal spectra are symmetric around their maximum,
 a consequence of the shape of the height distribution, which is symmetric around its maximum.
The symmetry of the $f(\alpha)$ spectrum, explicitly, is 
\begin{equation}
	\alpha_{q} + \alpha_{-q} = 2 \alpha_{q = 0}
\end{equation}
This relation is distinct from the symmetry relation obeyed by the multifractal spectrum of a disordered system at the Anderson transition \cite{Romer}: in the present case, the symmetry relates $q$ to $-q$, whereas the symmetry at the Anderson transition connects $q$ to $1-q$.
However, in both cases, the scaling of the \emph{small wavefunction components} (given by $q \to \infty$) is related to the scaling of the \emph{large wavefunction components} (given by $q \to - \infty$). As we remarked earlier, under the substitution $S'$ (\ref{eq:ArrowSub}) the heights on the ``old" sites $i$ change sign, and $\psi(i)$ is transformed to  $\psi(i)^{-1}$. In other words large components are transformed to small components and vice-versa.

\subsection{Generalization to other aperiodic chains}
One may ask to what extent the results of the preceding section are generically true for quasiperiodic chains. 
To investigate this question, we consider models on a series of quasiperiodic chains: the so-called metallic mean substitutions (based on the golden mean $\tau$ and its generalizations to silver, bronze, \dots irrationals) \cite{Thiem2011}.
They are generalization of the Fibonacci substitution:
\begin{equation}
\label{eq:subn}
	\sub_n: 
	\begin{cases}
		a \to a^n b\\
		b \to a,
	\end{cases}
\end{equation}
where $a^n$ is a shorthand notation for $a$ repeated $n$ times.
$n = 1$ yields the Fibonacci chain, $n=2$ the silver-mean chain, $n=3$ the bronze-mean chain, \dots

In this case, it is easy to show by direct inspection of the inflation rule for the arrows that the evolution of the height statistics still obeys the Fokker-Planck-like equation \eqref{eq:FP}.
Thus, the same conclusions hold as in the Fibonacci case: the partition function has the scaling behavior
\begin{equation}
	Z^{(2t)}_\mu(\beta) \simop{t \to \infty} \omega_n(\beta)^t f_\mu(\beta),
\end{equation}
the height distribution converges to a normal distribution, and the central state is multifractal with the fractal dimensions
\begin{equation}
	d_q(\kappa, n) = \frac{1}{q-1} \log \left( \frac{\omega_n(2 \kappa)^q}{\omega_n(2 \kappa q)} \right)/\log(\omega_n(0))
\end{equation}

For $n$ even, $n=2k$, we can compute the $\omega$ function explicitly:
\begin{equation}
	\omega_{2k}(\beta) = \frac{2 e^\beta + k^2 (1 + e^\beta)^2 + k (1 + e^\beta) \sqrt{4 e^\beta + k^2 (1 + e^\beta)^2}}{2 e^\beta},
\end{equation}
and thus find the explicit expression of the fractal dimensions of all the central states in the series $n = 2k$.
For $n$ odd, the substitution rule for the arrows is more involved and we were not able to compute explicitly the $\omega$ function.

\begin{figure}
	\centering
	\includegraphics[scale=0.9]{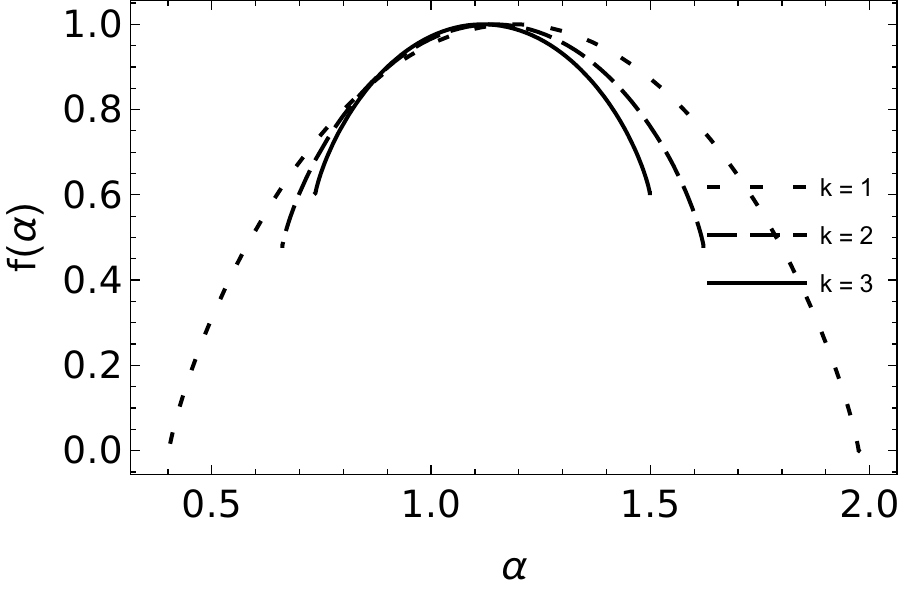}
	\caption{The multifractal spectrum of the central state of three metallic mean chains ($k=1,2,3$), for $\rho=1/2$.}
	\label{fig:6_multifractal_spectrum_k}
\end{figure}

Going back to the even case, one can readily check that, as $n$ grows, the central state behaves more and more like an extended state: $d_q(\kappa, n \to \infty) = 1$.
This is to be expected since the aperiodic chain locally looks more and more periodic as $n$ grows.
Figure \eqref{fig:multifractal_spectrum_k} shows the multifractal spectrum for different values of $k$. 
We see that the support of the spectrum decreases as $k$ is increased, which is yet another signature that the state becomes less critical as $k$ is increased.

To conclude this section, we consider an aperiodic substitution which is \emph{not} quasiperiodic:
\begin{equation}
\label{eq:B3Sub}
	\sub: 
	\begin{cases}
		a \to a b b b \\
		b \to a,
	\end{cases}
\end{equation}
This substitution, which we will call the \emph{b3 substitution}, 
is aperiodic -- but not quasiperiodic. Indeed, as the eigenvalues of the substitution matrix do not satisfy the Pisot condition \footnote{To be Pisot, only one of the eigenvalues can be greater than 1 in absolute value} the structure does not have Bragg peaks in its diffraction spectrum \cite{godrecheluck}. The b3 substitution has been thoroughly studied from the geometrical point of view, see eg. \cite{Frank}.
\begin{figure}
	\centering
	\includegraphics[scale=0.6]{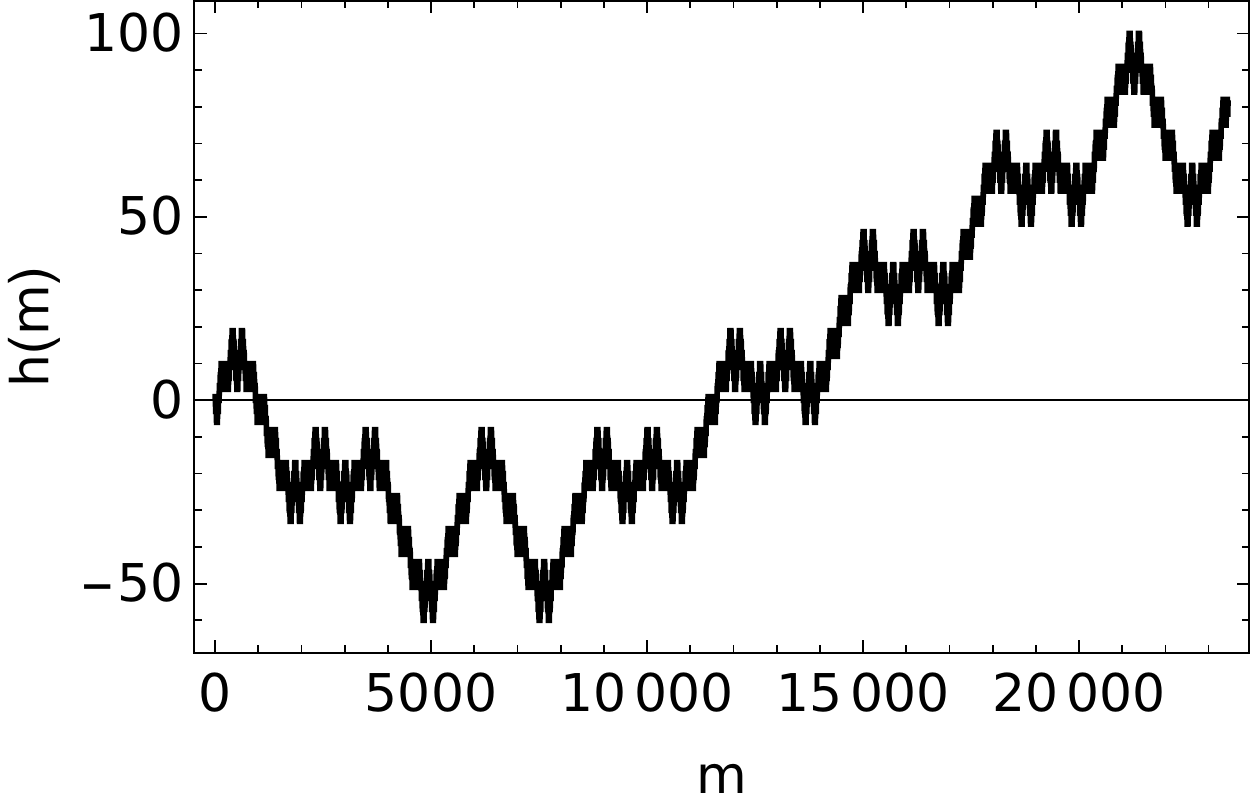}
	\caption{The field of heights for the aperiodic and non-quasiperiodic b3 substitution. The central state is localized, like in the case of a random chain, although this chain is deterministic.}
	\label{fig:height_b3}
\end{figure}

One can show that under the b3 substitution, the heights are multiplied by 3, along with a sign change, giving:
\begin{equation}
	\mathbf{N}^{(t+1)}(-3h) = \sum_{h' = -3}^{3} M(h') \mathbf{N}^{(t)}(h-h')
\end{equation} 
This equation implies that the height on this tiling (fig.~\eqref{fig:height_b3}) grow much faster than in the quasiperiodic case: in particular, the maximum of the height grows as a power law $h_\text{max}(L) = \max_{i \leq L} h(i) \sim L^\alpha$, whereas in the quasiperiodic case it grows much slower -- logarithmically.
The power-law exponent can be computed explicitly: $\alpha = \log 3 / \log \omega_\text{b3}^3$, where $\omega_\text{b3} = (1+\sqrt{13})/2$ is the largest eigenvalue of the b3 inflation matrix.
Since $h_\text{max}(L)$ behaves as a power-law, we conjectured and verified numerically that the typical height $h_\text{typ}(L)$ also behaves as a power-law, implying that the state has a stretched exponential form, and is therefore localized, as in the random case \cite{Offdiagdisorder}.

To conclude, we have seen that the central state of a tight-binding Hamiltonian on aperiodic chains can be described in terms of a height field, which is the integral of a field of arrows drawn on the chain.
The structure of the central state (as read from its fractal dimensions) is directly related to the geometrical properties of the height field.
In the metallic mean case, the height field grows slowly with the distance
	$h(L) \sim \sqrt{\log L}$,
which results in the state being critical.
We believe that this behavior is likely to generalize to canonical cut-and-project chains, of which the metallic mean chains are particular cases.
Next we considered an aperiodic chain which is \emph{not} quasiperiodic, the  b3 chain, whose behavior is seen to be radically different. the typical height now grows much faster,
	$h(L) \sim L^\alpha$,
and this results in the central state being localized. Note that, although this state has the same scaling characteristics as that on the random chain, this chain is purely deterministic.

\subsection{Transmission coefficient}

\begin{figure}
	\centering
	\includegraphics[scale=0.6]{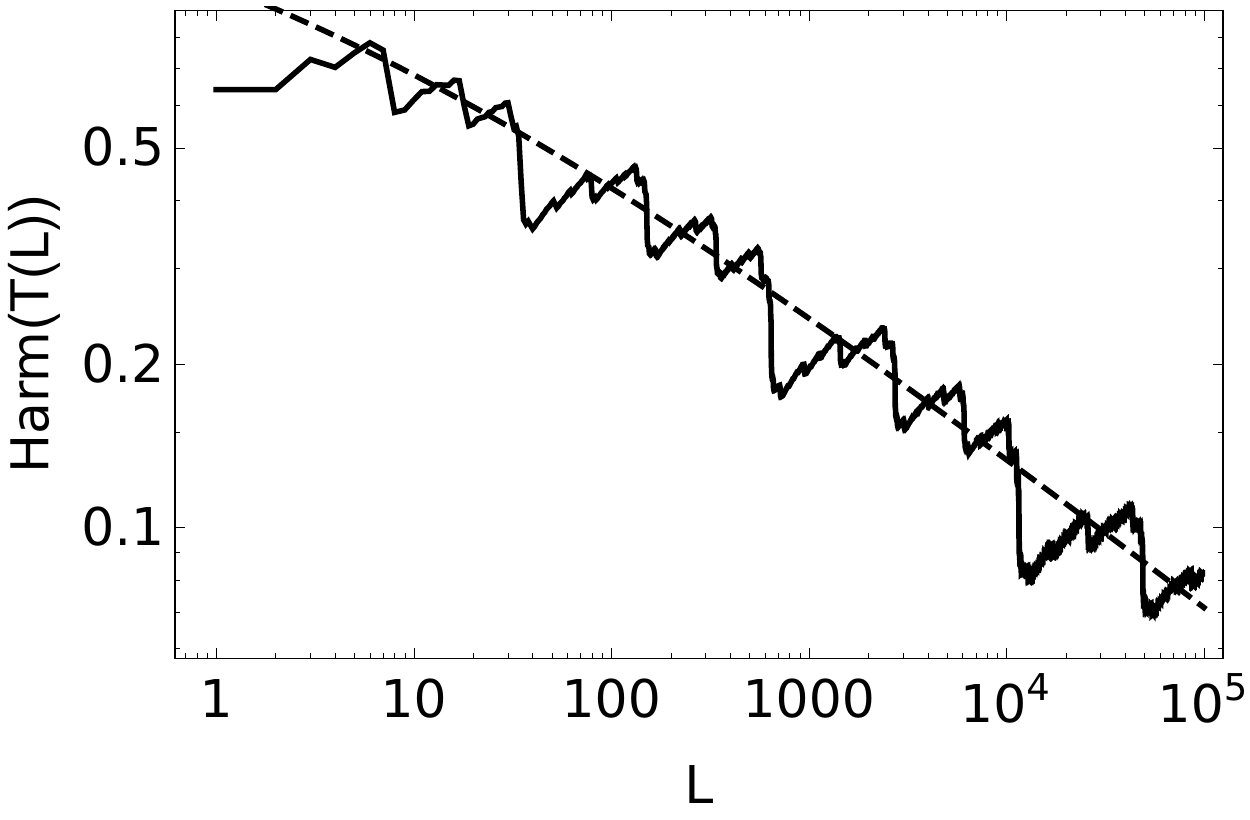}
	\caption{Mean (harmonic) transmission of a finite-size piece of length $L$ of the Fibonacci chain. Dashed line: analytical prediction \eqref{eq:mean_transmission}, continuous line: numerical calculation (average was performed on subchains of a system of $F_{27} = 196418$ atoms).}
	\label{fig:mean_transmission}
\end{figure}

We conclude this section with some results for the transmission coefficient, as an example of  the special properties of these critical states. 
Consider an approximant of the Fibonacci chain consisting of $2n$ atoms, attached to translationally invariant left and right leads.
The transmission coefficient is the fraction of the incident electronic probability that goes through the system.
Creating a plane wave at energy $E=0$ (i.e of wavevector $k = \pm \pi/2$) on the periodic lead excites the $E=0$ state of the chain. 
The transmission coefficient $T_n$ is then \cite{Beenakker}:
\begin{equation}
	T_n =\frac{4}{\left( x_n + \frac{1}{x_n} \right)^2}
\end{equation}
with $x_n = |\psi(2n)/\psi(0)|$.
Note that this transmission coefficient is proportional to the zero-temperature dc conductance of the system at Fermi energy $E=0$ \cite{conductance}. 

Exploiting the \SKK\ form of the $E=0$ state, $|\psi(2n)| = e^{\kappa h(n)}$, the transmission can be recast as
\begin{equation}
	T_n = \frac{1}{\cosh^2[\kappa (h(n)-h(0))]} ,
\end{equation}
so that the transmission between two sites depends only on the height difference of these two sites.
The transmission is maximal when the height difference is zero, in other words, {\it one can have perfect transmission between two sites of the chain which are arbitrarily far apart.} 

Another useful quantity which is easy to compute is the harmonic mean of $T$ defined by 
\begin{equation}
\label{eq:transmission}
	\langle T \rangle(L) = \left( \frac{1}{L}\sum_{i \leq L} \frac{1}{T_i} \right)^{-1}
\end{equation}
To estimate the typical transmission, the harmonic mean is preferred over the arithmetic mean in cases when the distribution is very wide leading to domination of the latter by rare events. 
This was pointed out for random systems \cite{conductance}, where it is the logarithm of $T$ which is distributed normally. 
Here too, the distribution of $\log T$ is expected to be Gaussian, as can be inferred from Eq.~\ref{eq:transmission}. 
The scaling of $\langle T \rangle(L)$ with system size can be calculated using the properties of the distribution of heights. 
The mean transmission of this chain after $t$ inflations is given by
\begin{equation}
\label{eq:mean_transmission}
\langle\text{T}\rangle(L)
    \sim 2 \left( 1 + (L/L_0)^{\log \frac{\om(2\kappa)}{\om(0)}/\log \tau^6} \right)^{-1}
\end{equation}
Since $\om(\kappa) \leq \om(0)$, the mean transmission goes to zero at large distance as a power law, in good accord with numerical data as seen in \eqref{fig:mean_transmission}.
Remark that the analytical calculation is performed for chains of even length, ie, every third approximant of the Fibonacci series, whose lengths are given by $L_t = F_{3t} \sim \ \tau^{3t}$.
This explains why the analytical calculation only captures the overall power-law behavior of the transmission, but not the superimposed log-periodic oscillations observed in the numerical study, which was carried out for all lengths $L$ (fig.~\eqref{fig:mean_transmission}).


\section{\SKK\ states on 2D tilings and their properties}
\label{sec:2D}

This section deals with the extension of the preceding calculations to two dimensional models. We will focus on the two most studied tilings: 
the Penrose rhombus tiling and the Ammann-Beenker tiling.

\begin{figure}
\centering
\includegraphics[scale=0.35]{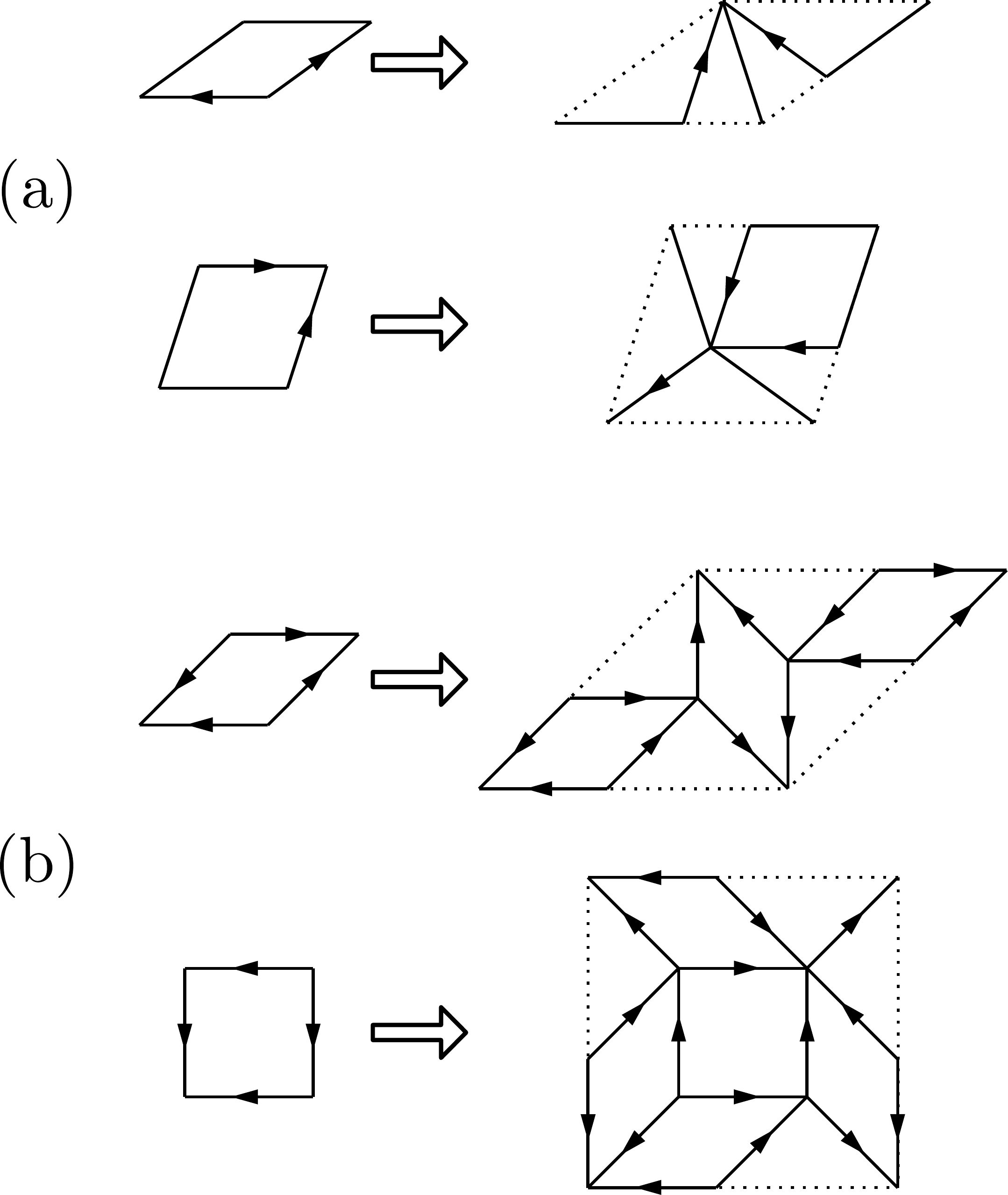}
\caption{ Tile substitution rules and arrow decorations corresponding to matching rules (see text) for (a) Penrose tiling and  (b) \AB\ tiling}
\label{fig:Infl2D}
\end{figure}

Both tilings are built using two tiles: these are the fat and thin rhombuses in the case of the Penrose tiling, and the square and 45$^\circ$ rhombus for the \AB\ tiling.
These tilings can be constructed by successive transformation of the two tiles using well-known substitution rules, as shown in fig.~\eqref{fig:Infl2D}. Furthermore, it is known that substitution tilings possess matching rules \cite{Goodman-Strauss}. When building a tiling by placing tiles one after another, these rules help to distinguish ``allowed" from "forbidden" configurations of adjacent tiles \cite{Baake2013}. 
In the case of Penrose and \AB\ tilings, the matching rules can be represented by arrows decorating the tiles. Arrows must point in the same direction for two tiles to be adjacent. Some arrows for tiles, and their inflations, are shown in fig.~\eqref{fig:Infl2D} for the Penrose and \AB\ tilings.

As can be checked by inspection, these arrows form an irrotational field.
This implies that they can be written as a gradient of a scalar function, the height function $h(i)$, taken up in Sec.~IIB. As in the case of 1D chains, the height functions will enter in the construction of \SKK\ states on 2D tilings. 

\subsection{Tight-binding models on 2D tilings}

We can now ask the question whether tight-binding models for electrons hopping along vertices of these 2D tilings admit eigenstates which can be defined in terms of the arrow field.

Sutherland was the first to answer positively to this question, furnishing in \cite{Sutherland} a tight-binding Hamiltonian in which on-site potentials are tuned to have specific values, so that its ground state is given by  $\psi(i) = \exp(\kappa h(i))$, where $\kappa$ is a constant. 
Note that this expression is not yet the form Eq.~\eqref{eq:SKK}), as it lacks the pre-exponential factor.
Similar eigenstates have been found by other authors for more complicated  Hamiltonians \cite{Tokihiro, Repetowicz}. 
In those works, a functional form is proposed, and used to reverse engineer the Hamiltonian for which it is an eigenstate.
These Sutherland-type solutions correspond to artificially ``fine-tuned'' Hamiltonians  and do not apply in the standard tight-binding case \eqref{eq:SimpleHam}. 
The solution for the ground state of the pure hopping Hamiltonian proposed by Kalugin and Katz \cite{KaluginKatz} was thus an important advance.

In this paper we consider quasiperiodic tight-binding Hamiltonians in 2D with off-diagonal as well as diagonal (onsite) terms, given by
\begin{equation}
\label{eq:GenHams}
	H(t, V) = -t \sum_{\langle i, j \rangle} c_j^\dagger c_i + \hc + V\sum_i z_i c_i^\dagger c_i 
\end{equation}
where $z_i$ is the coordination of site $i$, and where the hopping occurs between nearest neighbor sites. The parameter $V$ allows to go continuously from the pure hopping model ($V$=0) to the discrete Laplacian model ($V=t$) and eventually to the limit of decoupled atoms ($V/t \rightarrow \infty$). We will show that the ground state wavefunction  has the \SKK\ form for any choice of $V$, namely :
\begin{equation}
\label{eq:gswf}
	\psi_\text{GS}(i) = \loc(i) e^{\kappa h(i)}
\end{equation}
where $h(i)$ is a height field defined on the vertices of the tiling, and $\kappa$ is a constant. The pre-exponential factor $\loc(i)$ is, as we stated in the introduction, a quasiperiodic function which depends on the local environment of the site $i$.
This form of the wave function allows to compute exactly, as in 1D, the multifractal spectrum of the groundstate of this family of models. To do this, it is necessary to know the properties of the height field, studied in the next section.  

\subsection{Properties of the 2D height field. Fractal dimensions of the ground state.}

As for the one dimensional case, the height on a given site is the integral of an arrow field. 
The arrows correspond to decorations of edges of the basic tiles, as illustrated in Fig.~\eqref{fig:Infl2D} for the Penrose and \AB\ tilings.
Fig.~\eqref{fig:Infl2D} also indicates how the distribution of heights evolves under inflation.  
By repeated inflations, one thus generates the set of height fields for larger and larger pieces of tiling. 

\begin{figure}
\centering
\begin{subfigure}[b]{0.4\textwidth}
	\includegraphics[width=\textwidth]{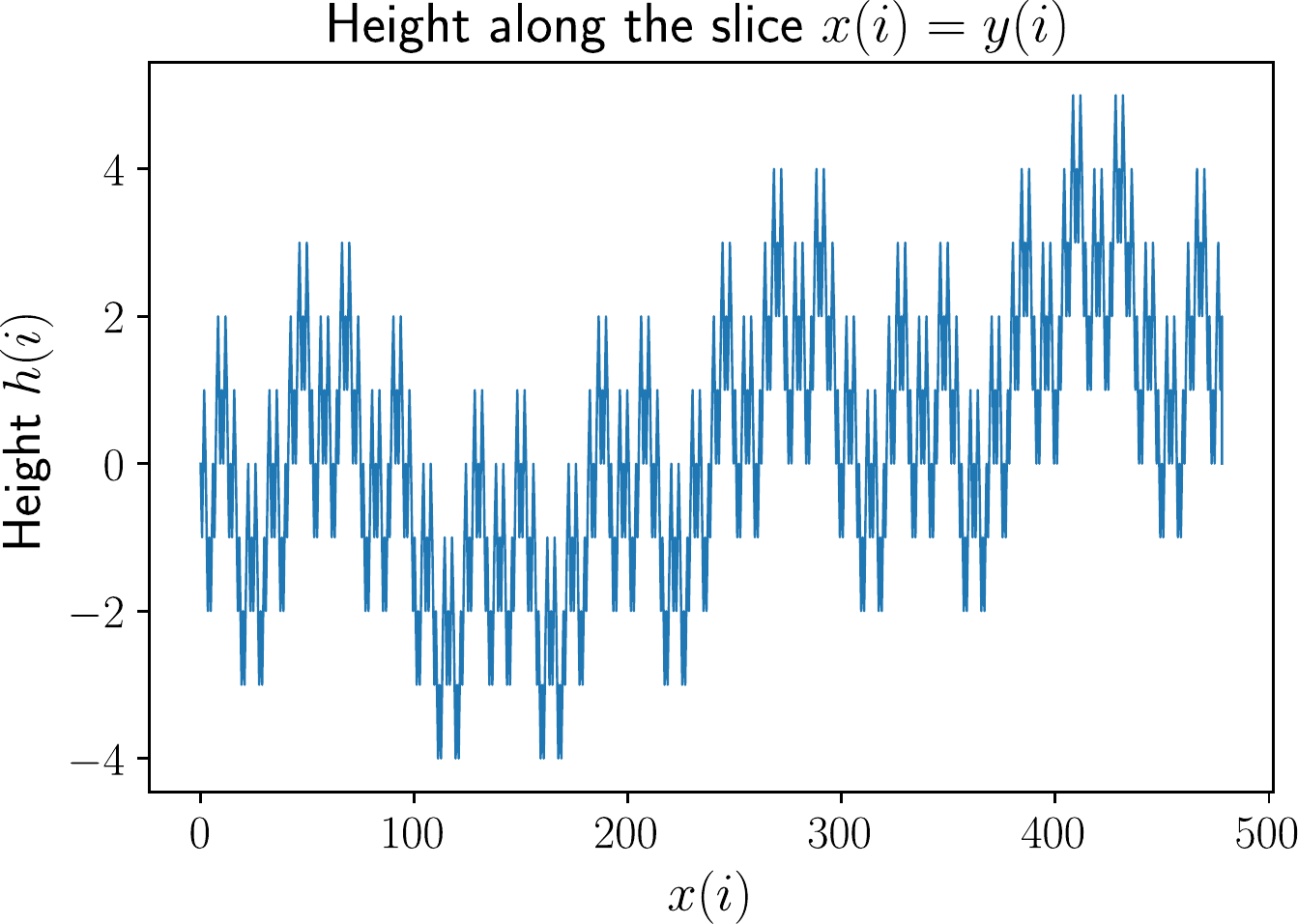}
    \caption{}
    \label{subfig:height}
\end{subfigure}
\begin{subfigure}[b]{0.5\textwidth}
	\includegraphics[width=\textwidth]{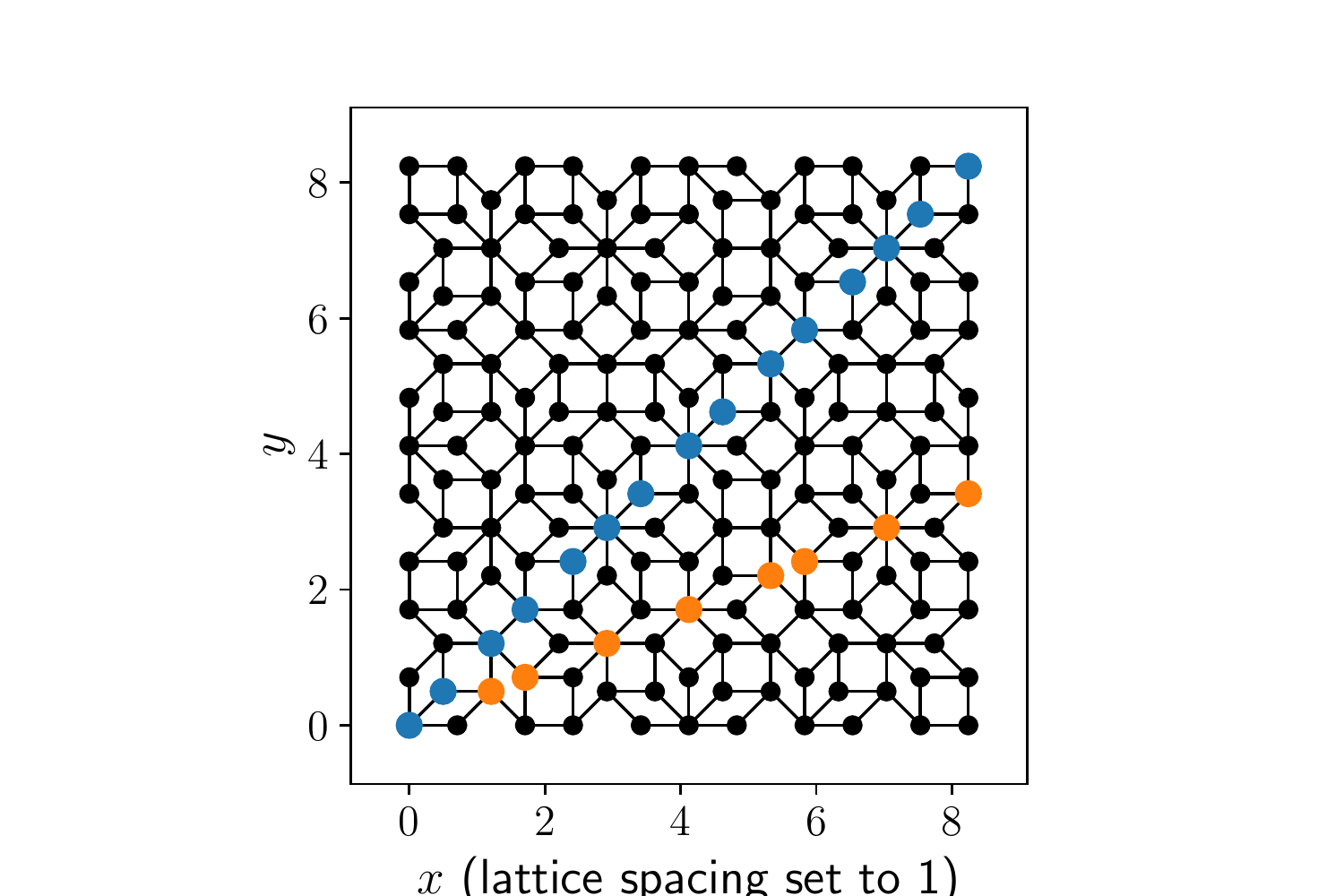}
    \caption{}
    \label{subfig:highlight}
\end{subfigure}%
\caption{(a): The height field along a slice of the \AB\ tiling. $i_x$ and $i_y$ is a shorthand notation for the $x$ and $y$ positions of the node $i$. (b): In blue, the slice taken in (a), and in orange another slice, along which the height is constant.}
\label{fig:height}
\end{figure}

To understand the spatial dependence of the height field, it is instructive to start by examining the variations of $h(i)$ along some specific directions in the tiling.
Fig.~\eqref{subfig:height} shows the height field along the slice of the \AB\ tiling shown in blue in figure \eqref{subfig:highlight}.
The similitude of behavior with the 1D height function of the Fibonacci chain (fig.~ \eqref{fig:arrows_heights}) is evident. 
Let us now explain the reasons for this similarity.

Consider a line which runs parallel to one of the eight possible orientations of the edges (in our figures these occur at angles which are multiples of $\pi/4$).
In fig.~\eqref{subfig:highlight} the blue line is shown as an example.
The distance between a pair of nodes along such a line is $n l' + m L'$, where $l'$ and $L'$ are respectively the length of an edge and of the diagonal of a square, and where $n$ and $m$ are integers.
Upon substitution \eqref{fig:Infl2D}, the lengths are transformed according to $l' \to l'+ L'$, $L' \to l'+ L'+ l'$. 
The lengths $A = l'+ L'$ and $B = L'$ are transformed according to the silver mean substitution: $A \to A + A + B$ and $B \to A$. 
Reasoning along the same lines as in sec.~\eqref{sec:1Dheight}, the heights along any such line oriented at angles $n\pi/4$ also transform according to the silver mean substitution, meaning that the height plot along such a line (fig.~\eqref{subfig:height}) should have the same structure as in the 1D case.

On the tiling, there are also paths along which the height is constant, as illustrated by the orange line shown on figure \eqref{subfig:highlight}, corresponding to directions lying in-between the axes of the edges (in our figures these occur at angles $(n+1/2) \pi/4$).
For the orange path, the distance between two nodes is of the form $n l + m L$, where $l$ and $L$ are respectively the length of the small and large diagonals of lozenges. 
Therefore, any two nodes along such a line are connected by a path going only through large and small diagonals of lozenges, a path along which, as can be seen from Fig.~\eqref{fig:Infl2D}, the height is a constant. 
This implies that the exponential factor of the wavefunction remains constant along these paths and any variations are due only to the prefactor $\loc$.
If one looks at the variation of the \SKK\ eigenstate along such a path, the state will appear to be extended, with no evidence of fractality. 

As in 1D, the scale invariance of the height field implies multifractality of the corresponding \SKK\ state. We present here the main results for the statistics of the heights, leaving the details for appendix~\eqref{sec:2Dheights}. 
The height statistics on the \AB\ and Penrose tilings obey the master equation
\begin{equation}
	\mathbf{N}^{(t+1)}(-h) = \sum_{h'} M(h') \mathbf{N}^{(t)}(h- h').
\end{equation}
The same conclusions hold as for the 1D case, namely : the height grows slowly  -- $h(L) \sim \sqrt{\log L}$, and the height distribution on a large tiling piece is Gaussian.
Moreover, the $\omega$ function -- which governs the large-scale behavior of the partition function -- can be computed exactly (see appendix~\eqref{sec:2Dheights}).
One finds for the \AB\ tiling
\begin{equation}
\label{eq:frac_ab}
    \om(\beta) = \frac{a(\beta) + \sqrt{a(\beta)^2 - e^{2 \beta}}}{e^\beta}
\end{equation}
with $a(\beta) = 4 \exp(2 \beta) + 9 \exp(\beta) + 4$.
In the case of the Penrose tiling we find
\begin{equation}
\label{eq:frac_penrose}
	\om(\beta) = \frac{b(\beta) + \sqrt{b(\beta)^2 - 4e^{2\beta}}}{2}
\end{equation}
with $b(\beta) = \exp(2\beta) + 5\exp(\beta) + 1$.

As in the 1D case, this $\om$ function determines the multifractal spectrum of the \SKK\ state $\psi$ through the equations :
\begin{equation}
\label{eq:multifractal}
	\alpha_q = \log(\omega(2 \kappa)) - 2\kappa \frac{\omega'(2q\kappa)}{\omega(2q\kappa)}
\end{equation}
and
\begin{equation}
	f(\alpha_q) = \log(\omega(2q\kappa)) - 2q\kappa \frac{\omega'(2q\kappa)}{\omega(2q\kappa)}.
\end{equation}
These theoretical predictions will be compared with results from numerical computations, as described in the next subsections.

\subsection{Results for the AB tiling}

The tight-binding matrix was diagonalized numerically for a fixed value of the onsite potential $V$ and for a range of system sizes.
In numerical computations, it is rather common to consider ``canonical approximants'': periodic tilings with arbitrarily large unit cells approximating the quasiperiodic tiling (see e.g.~\cite{Duneau}).
However on these canonical approximants, the height field is not single-valued. 
We therefore use the modified approximants -- with mirror boundary conditions instead of periodic boundary conditions -- proposed in \cite{KaluginKatz}, which do not have this drawback.
We shall label approximants of increasing size with an integer, $n$, which is equal to the number of inflations \eqref{fig:Infl2D} performed to build it.
Using these approximants, the values of the constant $\kappa$ and the pre-exponential factors of the expression Eq.~\eqref{eq:SKK} can then be determined.

\subsubsection{The constant $\kappa$}
The first step of the analysis consists of determining the unknown constant $\kappa$ in the expression for the ground state $\psi_\text{GS}$ from the numerical solution of $\psi$. 
This is done by the method used in \cite{KaluginKatz}, which relies on the following observation: if two sites $i_1$, $i_2$ of the approximant have nearly identical local environments, then their wavefunction amplitudes will have almost identical prefactors.
The ratio of wave function amplitudes on these two sites is then 
\begin{equation}
	\frac{ \psi_\text{GS}(i_1) } { \psi_\text{GS}(i_2) } = \frac{ \loc(i_1) } { \loc(i_2) } e^{\kappa ( h(i_1) - h(i_2) )} \simeq  e^{\kappa ( h(i_1) - h(i_2) )}
\end{equation}
This relation allows us to calculate a value of $\kappa$ from the numerical solution, since the height field is known exactly for the approximant. One expects this value to become progressively more accurate as larger and larger approximants are considered, since the numerical solution approaches the exact one, and since one can find sites $i_1$ and $i_2$ whose environments are similar out to a greater distance. 
As can be seen in Tab.~\eqref{tab:kappa}, the $\kappa$-value determined in this fashion indeed converges. 

\begin{table}[htp]
  \begin{tabular}{ | c |  c  c  c | }
      \hline
      &&Penrose tiling&\\
      \hline
      $N$ & 9045 & 23490 & 61191 \\ \hline
      $\lambda_p$ & 0.02125888 & 0.02163468 & 0.02148895 \\ \hline
      && Ammann-Beenker tiling& \\
      \hline
      $N$ & 4180 & 23950 & 138601 \\ \hline
      $\kappa$ & 0.61215 \qquad& 0.61252 \qquad& 0.61249 \\
      \hline 
  \end{tabular}
  \caption{ $\kappa$ for different values of $N$ (the number of sites) in Penrose and \AB\ approximants for $V = -0.5$}
   \label{tab:kappa}
\end{table}
\begin{figure}
\includegraphics[width=0.4\textwidth]{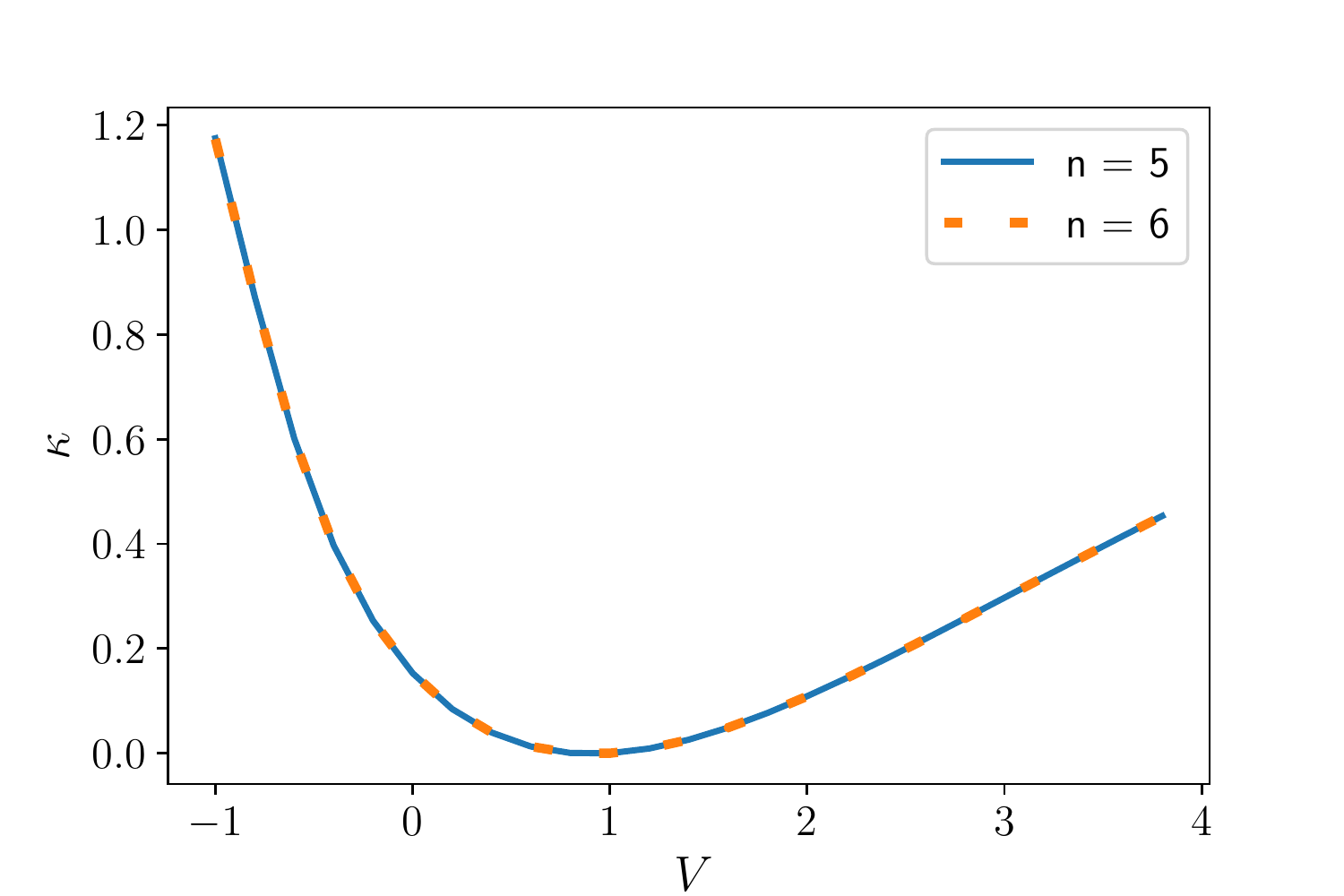}
\caption{ Plot of $\kappa$ versus $V$ for the ground state of the \AB\ tiling computed by numerical diagonalization for approximants $n=5$ and 6. }
\label{fig:kappa}
\end{figure}

Figure \eqref{fig:kappa} shows the evolution of $\kappa$ when the strength of the on-site potential, $V$, is varied in the model. The shift of the two curves in going from the $n=5$ to $n=6$ approximant is too small to be visible on the graph, indicating that the results are converged.

\subsubsection{The prefactors $\loc$}
\label{sec:prefactor}
\begin{figure}
\centering
\begin{subfigure}[b]{0.35\textwidth}
	\includegraphics[width=\textwidth]{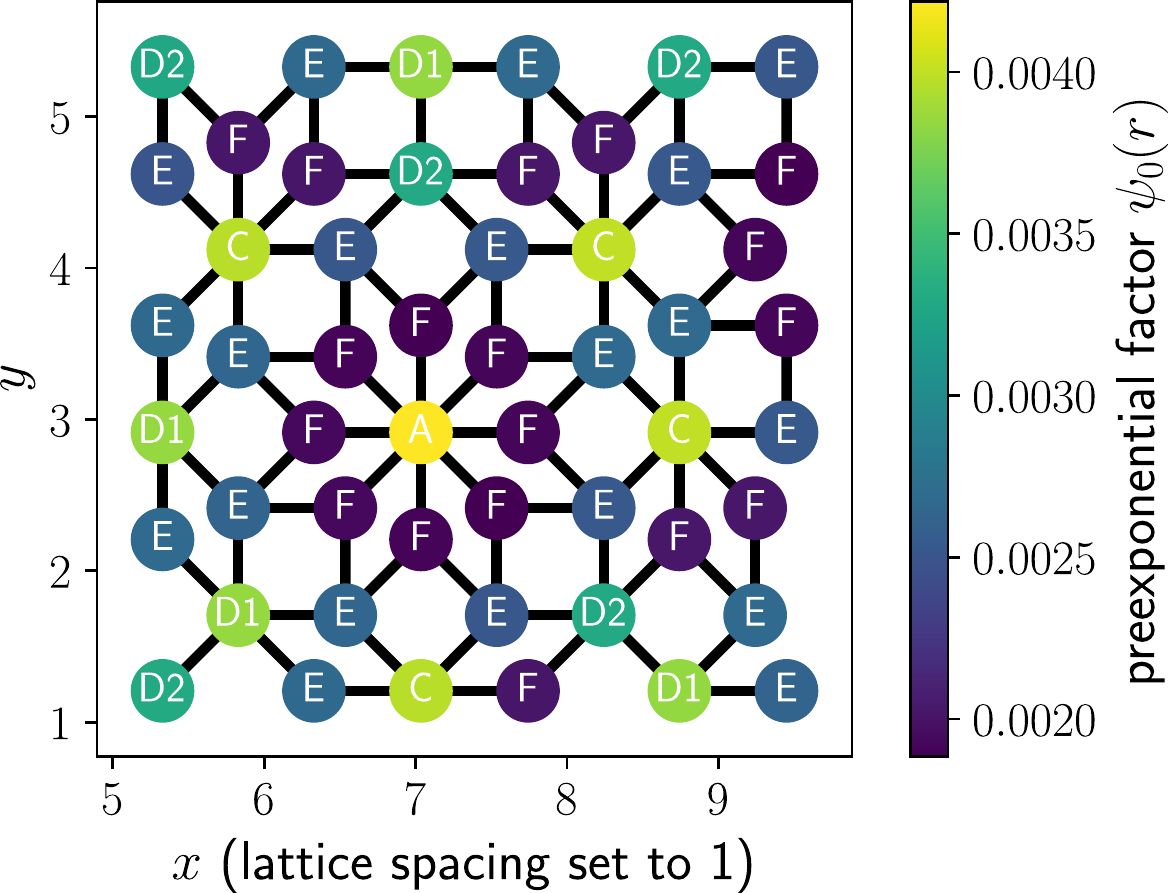}
    \caption{The numerically computed prefactors on a small patch of the approximant  $n=7$ (138601 atoms). }
    \label{subfig:subgraph}
\end{subfigure}
\begin{subfigure}[b]{0.4\textwidth}
	\includegraphics[width=\textwidth]{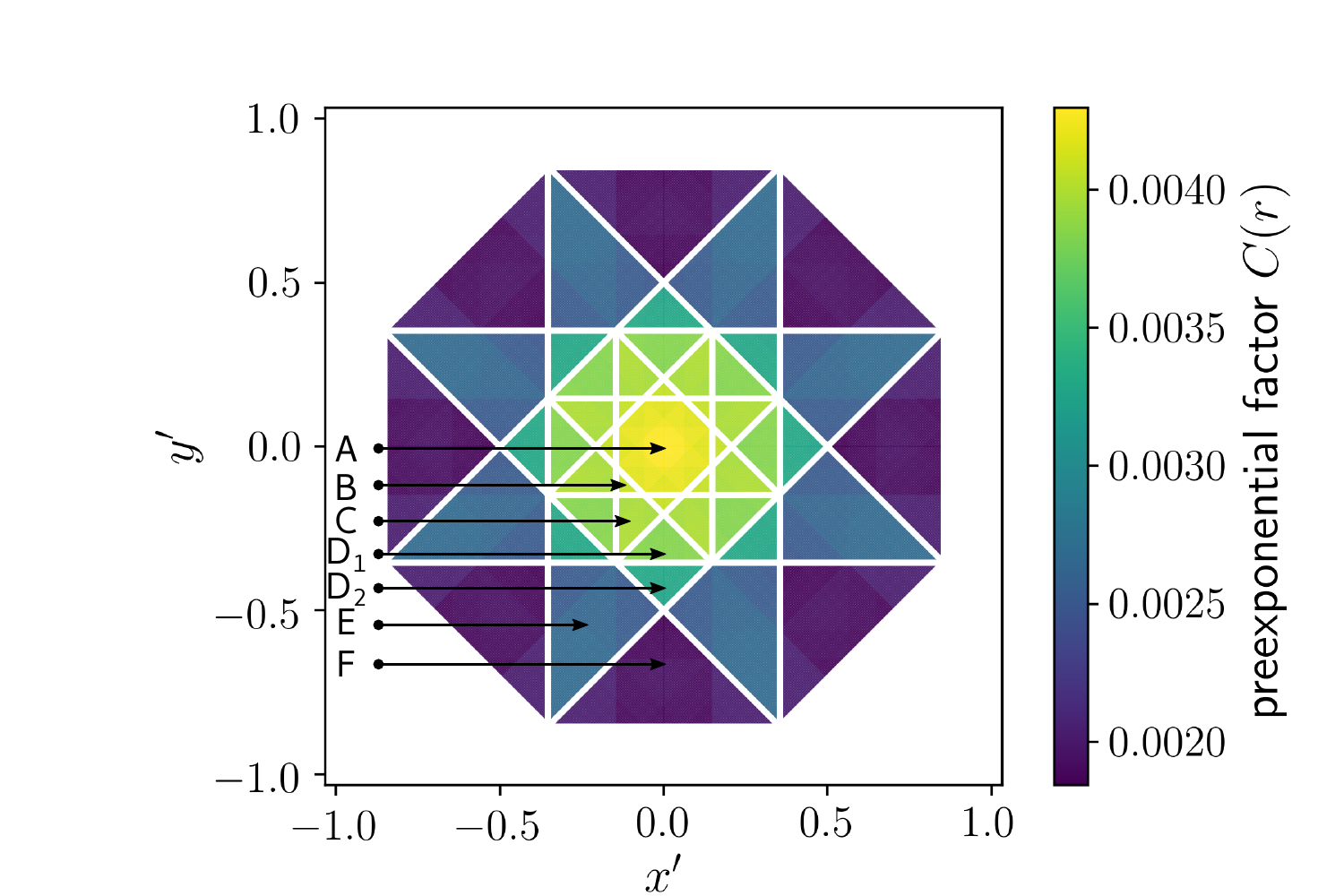}
    \caption{The numerically computed prefactor in perpendicular space, for approximant $n=7$. White lines outline the zones corresponding to different first neighbor environments.}
    \label{subfig:ABcake}
\end{subfigure}
\begin{subfigure}[b]{0.35\textwidth}
    \includegraphics[width=\textwidth]{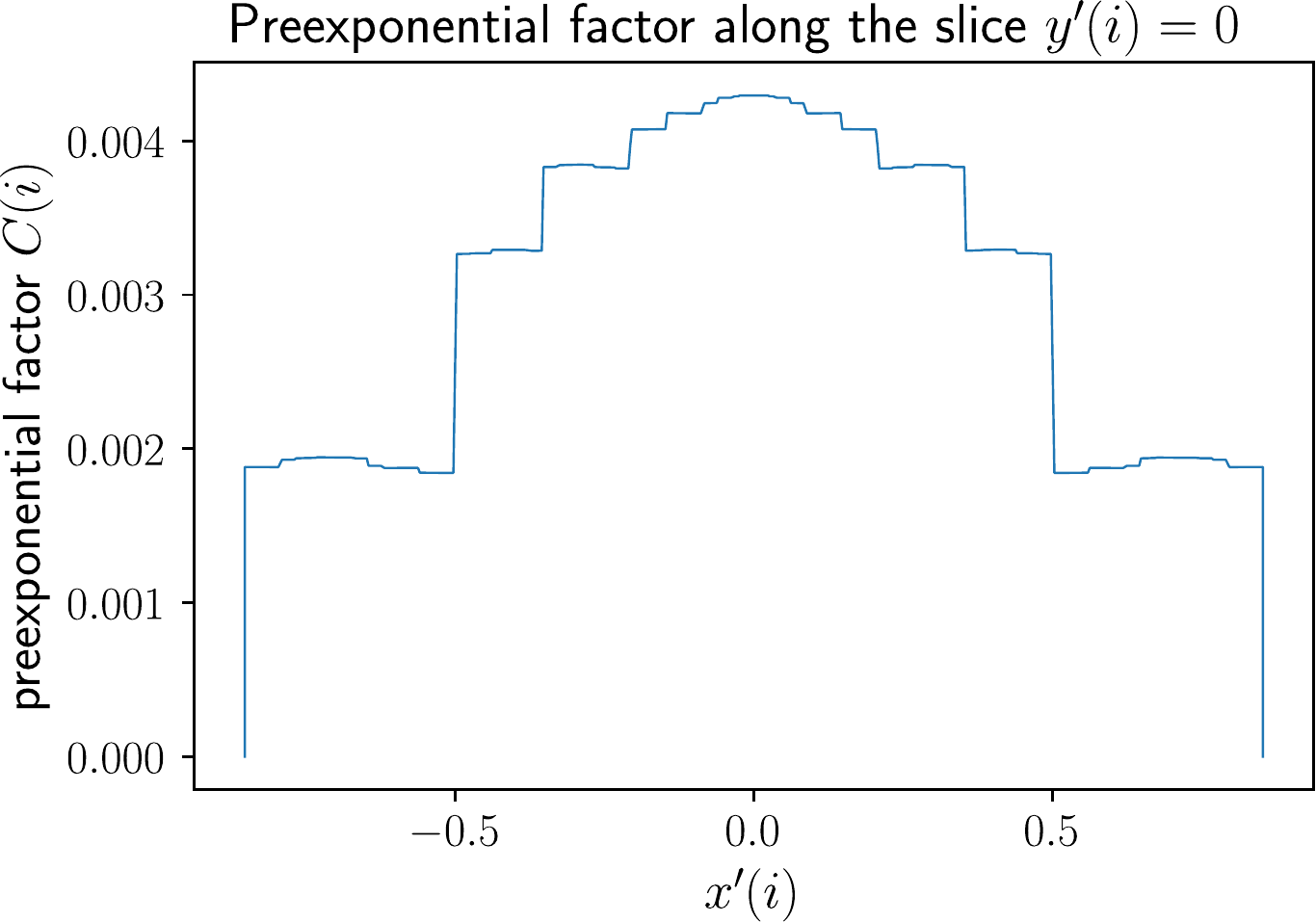}
    \caption{``Wedding cake" plot showing the variation of the prefactor along a horizontal cut passing through the origin in perpendicular space.}
    \label{subfig:ABcakeslice}
\end{subfigure}
\caption{}
\label{fig:preexp}
\end{figure}

The second step of the analysis consists of determining the set of prefactors $\loc(i)$ by plugging in the value of $\kappa$ already found. Once these have been been determined, we need to check whether $\loc(i)$ is local: that is, dependent only on the local arrangement of atoms. A first indication that this is indeed the case is provided by
Fig.~\eqref{subfig:subgraph}, which shows the numerically computed prefactors in a small region of the \AB\ tiling. The colors reflect the wavefunction amplitudes computed for the pure hopping Hamiltonian ($V = 0$), and sites have been labeled by A, B, C, D$_1$, D$_2$, E and F, to indicate the seven different nearest neighbor configurations on this tiling. A represents sites with 8 neighbors, B represents sites with 7 neighbors, and so on. The subscripts allow to differentiate between sites of the same coordination number (5), but which transform differently under substitutions. 
Upon inspection of the figure, the colors and labels are seen to correlate, providing a visual check that the pre-exponential factor depends primarily on the local arrangement of the atoms. 

A more quantitative demonstration of the local character of the $\loc(i)$ consists of plotting them in the perpendicular (or internal) space representation of the tiling. This representation allows to classify sites according to their environments on the tiling.  As detailed explanations can be found in the literature, it suffices here to give the general idea. Recall that the \AB\ tiling, as many other quasiperiodic tilings, can be constructed from a periodic tiling in higher dimension : the so-called cut-and-project construction method (see \cite{KaluginCP,DuneauCP,ElserCP,Baake2013} for a general introduction, and \cite{Duneau} for a description in the particular case of the \AB\ tiling).
In the case of the \AB\ tiling, the higher-dimensional space is 4 dimensional, and decomposes into the 2D physical plane, and the 2D orthogonal plane called the \textit{perpendicular (or internal) space}. 
The closer the projections of two vertices are in internal space the more similar their local environments in physical space
\footnote{More precisely, for any point $p$ and for any distance $r$, there is a $\epsilon(p,r) > 0$ such that any point closer than $\epsilon(p,r)$ from $p$ in internal space has a local environment that matches the one of $p$ up to distance $r$ in physical space.}. 
Internal space representation is thus well-suited to check the local character of the pre-exponential factor, as was already pointed out in \cite{KaluginKatz}.

Fig.~\eqref{subfig:ABcake} shows the $\loc(i)$ represented by a colorscale in perpendicular space. 
The seven families of nearest neighbor configurations A, B, \dots correspond to seven nonoverlapping domains as shown by the labels in the Fig.\ref{subfig:ABcake}. 
The colors and the domain labels are seen to be in good correspondence, showing thereby that the prefactors are determined by the local environment.

Within each family of environments, one can distinguish subfamilies according to the next nearest neighbor configurations. 
These in turn can be subdivided and so on. 
The prefactors $\loc(i)$, in consequence, exhibit a fine structure due to these differences of environments at the level of $n$th near-neighbors. This can be seen in Fig.~\eqref{subfig:ABcakeslice}, which shows the variation of the prefactor along a horizontal cut passing through the origin in internal space.

\begin{figure}
\centering
	\includegraphics[width=0.4\textwidth]{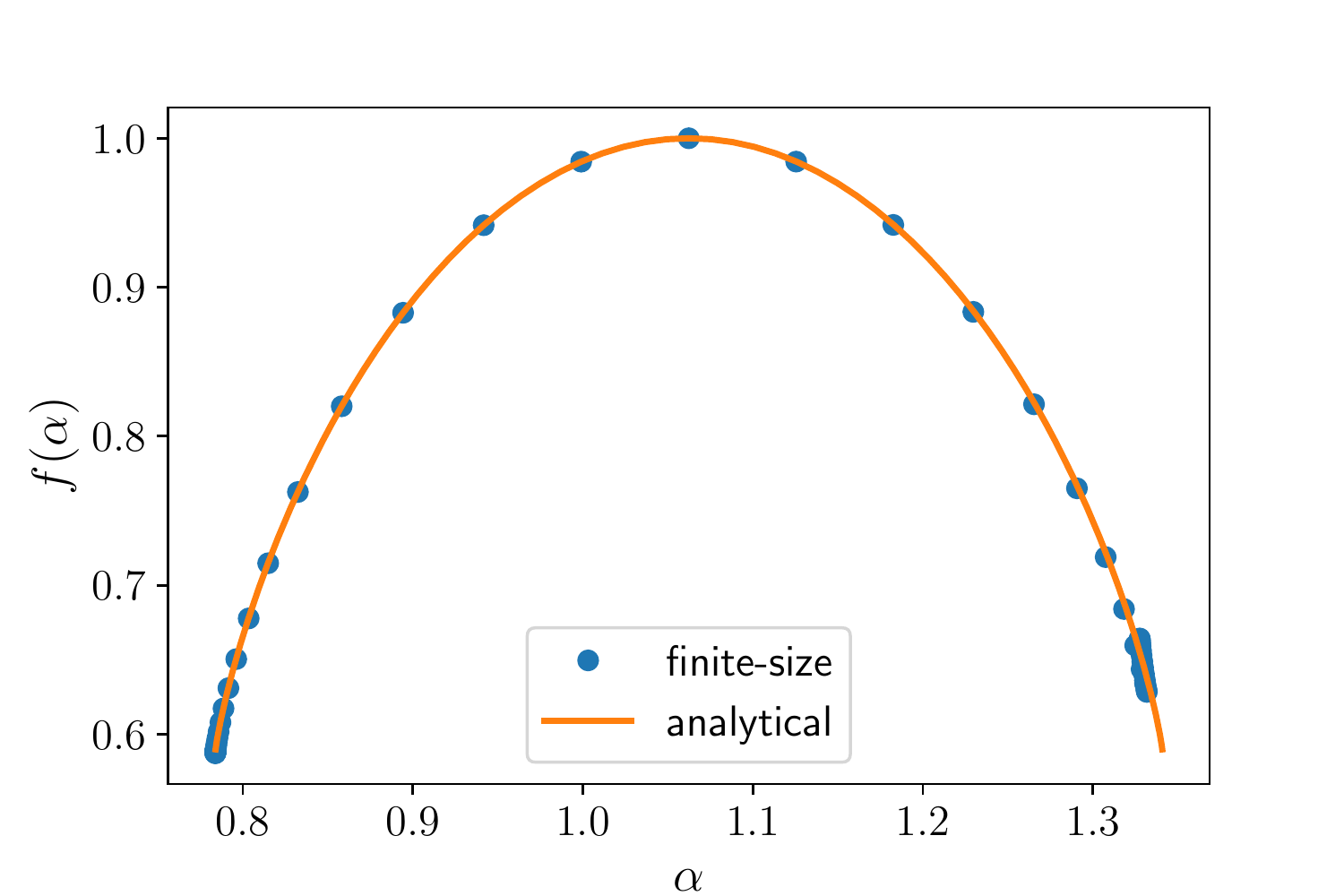}
	\caption{Dots: The numerically computed $f(\alpha)$ of the GS on the \AB\ tiling ($V=-0.5$). Solid line: the analytical solution \eqref{eq:multifractal}, using the numerically determined $\kappa$ (see text).}
\label{fig:multifractal}
\end{figure}

\subsubsection{Multifractal spectra}
 Fig.~\eqref{fig:multifractal} shows the multifractal spectrum, $f(\alpha)$, of the ground state calculated numerically for $\kappa(V = 0) \simeq 0.153$, on the \AB\ tiling. 
Dots indicate the  numerical values obtained from system-size scaling (see e.g.~\cite{Thiem} for a description of the method) for approximants up to $n=7$ (138601 atoms). 
One can see that the theoretical prediction and the numerical computations are in perfect agreement, showing that Eq.~\eqref{eq:multifractal} correctly captures the multifractal properties of the groundstate on the \AB\ tiling, for the family of models \eqref{eq:GenHams}.

The multifractality of a wavefunction can be measured by the width (or support) of the bell-shaped curve (see fig.~\eqref{fig:multifractal}) formed by its multifractal spectrum:
\begin{equation}
	\supp = \lim_{q \to \infty} \left( \alpha_{-q} - \alpha_q \right)
\end{equation}

For the \SKK\ states on the Penrose and \AB\ tilings
 we find
\begin{equation}
	\supp(\kappa) = \frac{4 |\kappa|}{\log \omega(0)}.
\end{equation}
Interestingly, this expression is valid not only for the Penrose and \AB\, but also for the Fibonacci chain, and for every chain of the metallic mean series, with $n$ even (see Sec. 2). 
For the metallic mean with $n=3$ however, the expression is different. 
To our knowledge, there is no explanation for the surprising universality of the support of the multifractal spectrum of the \SKK\ states.
The value of  $\kappa$, as shown on fig.~\eqref{fig:kappa} is proportional to the support of the multifractal spectrum, and thus tells us to which extent the groundstate is multifractal. To give a specific example, the figure shows that for $V=1$, $\kappa = 0$, and the groundstate is not multifractal at all -- it is extended. 
This is of course to be expected since for $V=1$ the model is just the discrete Laplacian, whose ground state is uniform over the tiling.
As $V$ is increased or decreased with respect to $V=1$,  $\kappa$ increases, and the support of the groundstate likewise. This state is thus becoming increasingly multifractal.

\subsection{A variational calculation for the ground state}
\label{sec:var}

\begin{figure}[htp]
	\centering
    \includegraphics[width=0.5\textwidth]{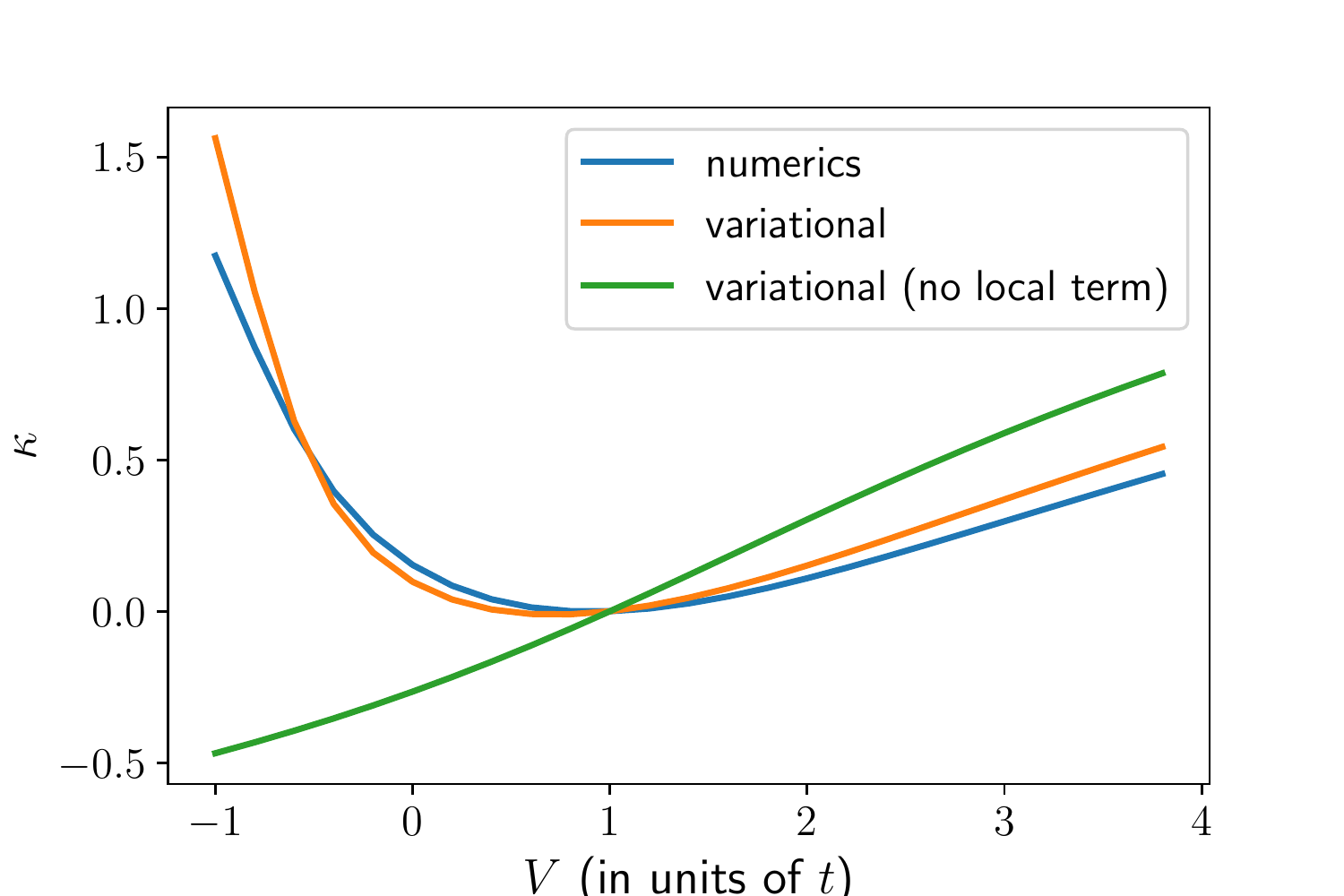}
    \caption{$\kappa$ for the model \eqref{eq:GenHams}. Blue: numerical result. Orange: prediction obtained using the variational method. Green: prediction of the variational method, for a variational state with no local term.}
    \label{fig:variational}
\end{figure}
It is interesting to ask at this point whether an analytical calculation could reproduce, even approximately, the \SKK\ solution just presented. In this subsection, we describe a variational calculation for the ground state on the \AB\ tiling. This method allows us to obtain an approximate theoretical solution for the ground state wave function and energy, and we will see that the results compare well with the numerical solutions.

Consider a variational ansatz which includes the exponential nonlocal term, and with a truncated set of local environments, namely the seven nearest neighbor environments present on this tiling. This approximation is expected to be reasonable, as can be seen from the numerical results shown in
(fig.~\eqref{fig:preexp}).  
The following form was thus chosen for the variational ansatz: 
\begin{equation}
\label{eq:variational}
	\psivar(i;\locvar, \kappa) = \locvar(i) e^{\kappa h(i)}
\end{equation}
where $\locvar$ is a function that takes seven values according to the seven possible first neighbor configurations of the \AB\ tiling (see fig.~\eqref{subfig:subgraph}).
The seven values of the function $\locvar$ and the parameter $\kappa$, are determined using a variational approach: $\locvar$ and $\kappa$ are chosen such that the functional
\begin{equation}
	E(\locvar, \kappa) = \frac{\bra{\psivar} H \ket{\psivar}}{\bra{\psivar}\ket{\psivar}}
\end{equation}
is minimized. Details of this calculation are given in the Appendix \ref{app:var}.

\begin{table}[htp]
\centering
\begin{tabular}{|c|c|c|c|c|}
\hline 
  &  ansatz 1 & ansatz 2  & ansatz 3 & Numerics \\ 
\hline 
$-E_\text{GS}$ &  4.16143 & 4.21936 & 4.22091 & 4.221697 \\ 
\hline 
$\kappa$   & -0.265337 & $\emptyset$  & 0.097175 & 0.153035 \\ 
\hline 
\end{tabular}
\label{tbl:gs_comp} 
\caption{Predictions for the ground state energy of the pure hopping model by the variational method.}
\end{table}

Figure \eqref{fig:variational} shows the value of $\kappa$ predicted by the variational method, together with its numerically determined value, as a function of the parameter $V$. 
To investigate the importance of the two factors entering the variational ansatz Eq.~\eqref{eq:variational}, we have given the results for variants as follows: ansatz 1 (all prefactors replaced by a constant), ansatz 2 (exponential terms replaced by a constant) and ansatz 3 (both factors allowed to vary). 
One sees that the agreement between the variational and the numerical values are quite good for ansatz 3. 
Increasing the number of environments considered in the variational ansatz would of course improve the quality of the agreement at the price of increasing the number of variables for minimization. 
 
It is interesting to compare the result of the variational method for ansatz 1 in which all prefactors are taken to be equal (the green curve in Figure~ \eqref{fig:variational}) with the ansatz 3, where prefactors are allowed to vary (the orange curve in Figure~ \eqref{fig:variational}), and with the numerical results.
One can see that, in the absence of the local prefactors, the $\kappa$ value is not predicted accurately.
To conclude, the variational method gives satisfactory values for $\kappa$ already at first-neighbor level of the approximation. 
We see that although the local term has no influence on the multifractal spectrum of the state -- whose shape is determined by $\kappa$ alone -- it must be included in the expression of the ground state in order to get a good value of $\kappa$.


\subsection{Results for the Penrose tiling}

\begin{figure}[htp]
	\includegraphics[width=0.4\textwidth]{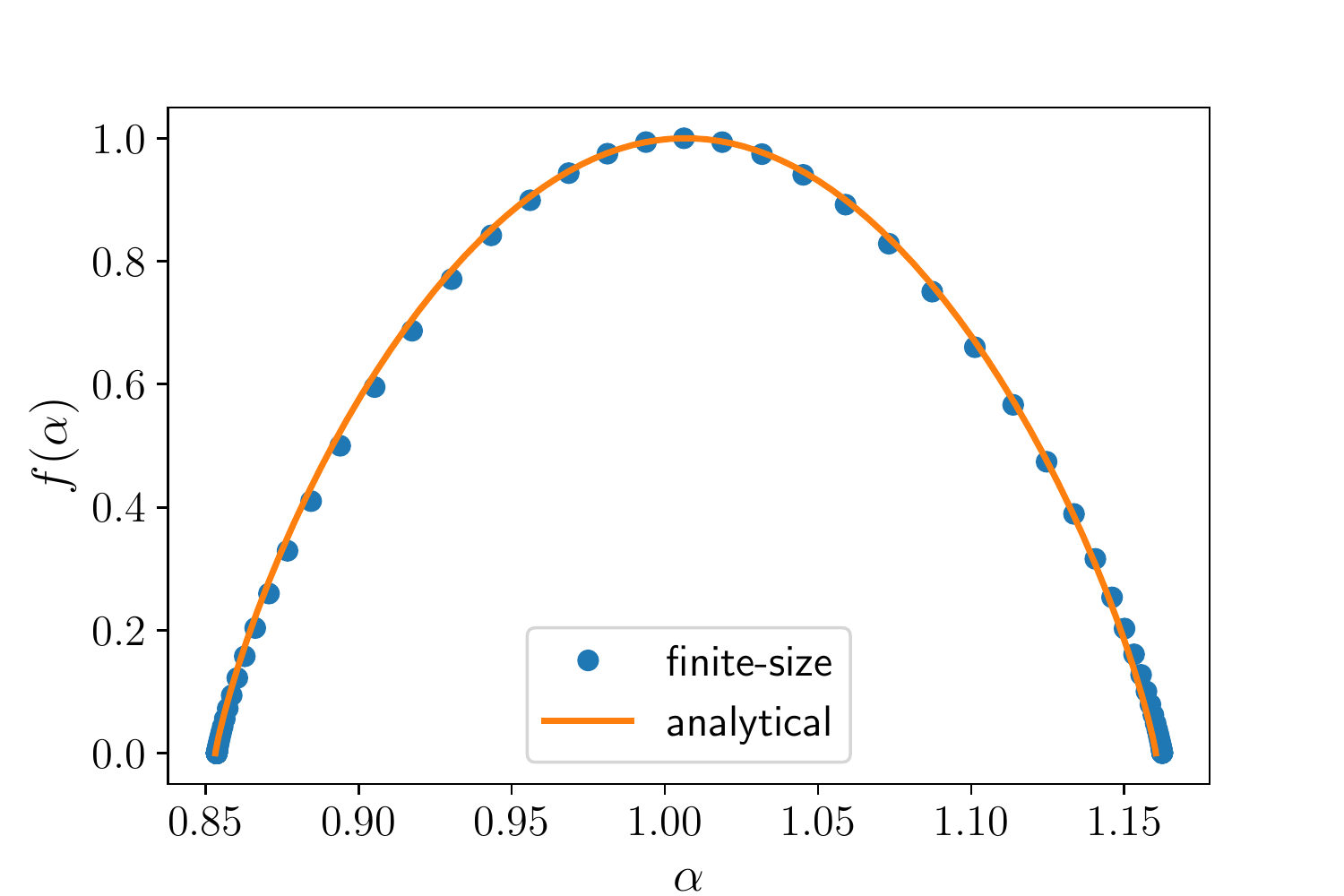}
    \caption{Dots: $f(\alpha)$ for the GS on the Penrose tiling ($V=-0.5$). Solid line: analytical curve \eqref{eq:multifractal} using numerically determined $\kappa$ (see text).}
\label{fig:Penrose}
\end{figure}

\begin{figure}[htp]
	\includegraphics[width=0.4\textwidth]{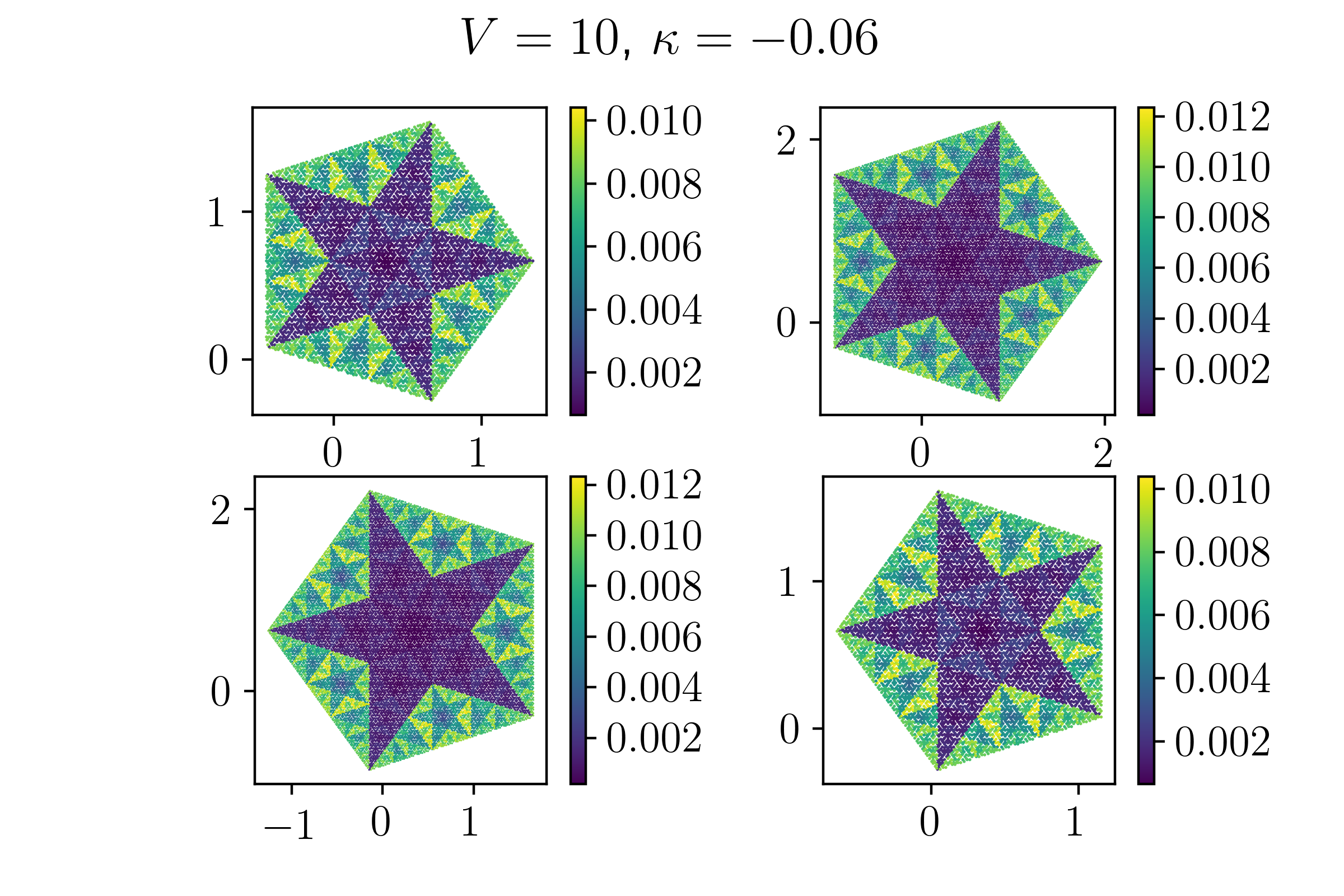}
    \caption{The prefactors of the ground state of \eqref{eq:GenHams} on the Penrose tiling represented in the four parallel planes of internal space. Here the approximant used is $n = 12$ (61191 atoms).}
     \label{subfig:Penrose_cake}
\end{figure}
\begin{figure}[htp]
	\includegraphics[width=0.4\textwidth]{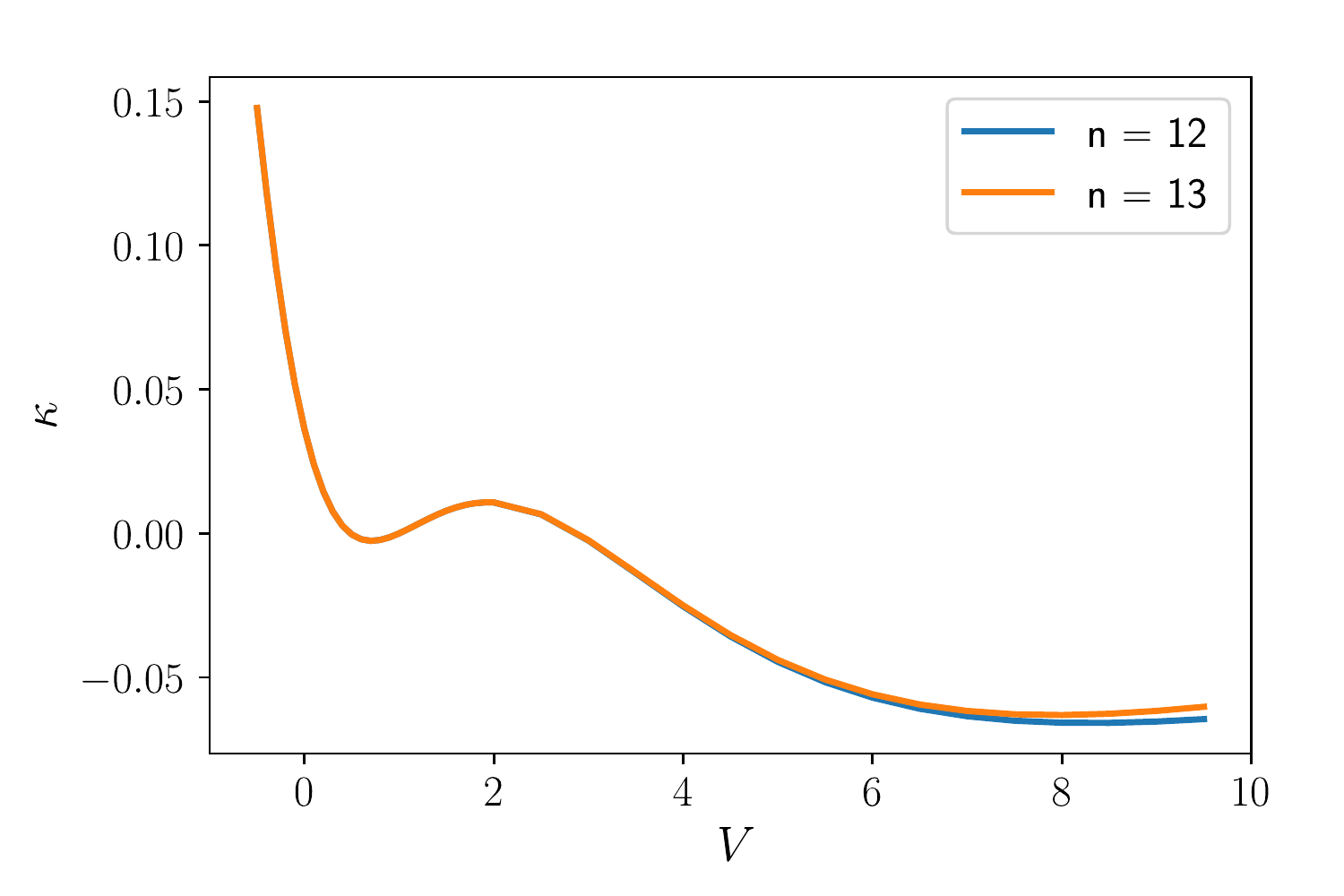}
    \caption{$\kappa$ versus $V$ for the Penrose tiling calculated for two approximants (61191 and 159705 sites).}
\end{figure}

In the case of the Penrose tiling, we proceed in the same way as described above for the \AB\ tiling. For arbitrary $V$, we verify that its ground state is of the \SKK\ type by computing the values of $\kappa$ and the pre-exponential factors. 

The evolution of $\kappa$ with $V$ is plotted in Fig.~\eqref{fig:Penrose}, where
one sees a significant difference between the two tilings. 
While the $\kappa(V)$ function for the \AB\ groundstate reaches 0 only at the trivial $V=1$ point -- meaning that the groundstate is critical and multifractal for all other values of $V$, the $\kappa(V)$ function for the Penrose groundstate has a zero at $V_0 \simeq 3$, a second non-trivial point at which the groundstate is extended. 
For $V=V_0$, the ground state amplitude depends only on the local environment, and therefore should be well approximated by a piecewise constant function over the (finitely many) k$^\text{th}$ nearest neighbor environments.

The pre-exponential factors are shown to be environment dependent by plotting them in internal space (fig.~\eqref{subfig:Penrose_cake}), as we did in the \AB\ case. 
For this tiling, the projections of vertices fall into four distinct pentagonal domains (corresponding to different planes of the 3D internal space), as are illustrated in fig.~\eqref{subfig:Penrose_cake}.

The fractal dimensions of the groundstate computed numerically are shown in Fig.~\eqref{fig:Penrose}. The curves corresponding to the two approximant sizes overlap, showing convergence of the results. The agreement with the analytical prediction eq.~\eqref{eq:multifractal} is likewise good.


\section{Discussion and Conclusions}

In this paper we have examined a family of tight-binding models on 1D and 2D quasiperiodic  tilings and shown that they admit solutions of the \SKK\ form: $\psi(i) = \loc(i) \exp(\kappa h(i))$. 
The prefactor $\loc$ depends only on the local arrangement of the atoms, while the exponential factor contains the height field $h$, which is nonlocal. It is this latter term which results in multifractality of these states.  We thus provide the theoretical demonstration of a property that had been numerically observed in the literature for eigenstates of quasiperiodic tilings.

In 1D, by considering the Fibonacci chain, we show that the states in the middle of the spectrum can be written in terms of an exponential of a height function. We described the properties of the height field and shown how it is linked to the geometry of the underlying quasiperiodic tiling. 
Extending the arguments to the metallic-mean family of quasiperiodic 1D chains, we show that the $E=0$ state is always multifractal. The multifractal spectrum for the $E=0$ state on the quasiperiodic chains is seen to have a reflection symmetry around its maximum for any value of the ratio of hopping amplitudes.
In contrast, we find that the $E=0$ state is localized for the case of the aperiodic $b3$ chain, which is not quasiperiodic. 

It is interesting to note that the ground state of these quasiperiodic 1D chains is {\it not} of the \SKK\ form. This state is not expressible as a product of an exponential function of a non-local heights function and a local prefactor.  Details of the demonstration can be found in Appendix D. 

We conclude the discussion of the 1D eigenstates by computing the expression for the exact transmission coefficient of a finite piece of the Fibonacci 1D chain when connected to perfect leads. The scaling of the typical transmission for a given length is also given. 

For the 2D Penrose and \AB\ tilings, we have studied a family of tight-binding models where the strength of the diagonal terms of $H$ can be continuously varied. We showed that the form of the ground state is preserved, while the constant $\kappa$ and the local prefactors vary. We computed analytically the properties of the height fields and fractal dimensions of the eigenstates, and compared the predictions to numerical results. We have discussed a variational calculation which allows to compute the 2D ground state wave functions to good approximation.

We conclude with some open questions. 
In the 1D case, we have constructed a height field by interpreting groups of two letters as arrows. One could do the same thing with groups of 3, 4, \dots letters. Whether the corresponding height fields describe eigenstates remains an open question.
Another point concerns extensions to other models and finding other 2D quasiperiodic tilings that host \SKK\ eigenstates. 
An interesting model to study in this context would be the Socolar tiling \cite{Socolar} with similarities to Penrose and \AB\ tiling and having 12-fold symmetry.
These questions remain for future study.

\textit{Acknowledgements.} 
We are grateful to Jean-No\"el Fuchs, Jean-Marc Luck and Julien Vidal for many fruitful discussions.

\appendix


\section{Fractal dimensions}

In the following, we consider an eigenstate of the form $\psi_i = \loc(i) e^{\kappa h(i)}$, on a quasiperiodic tiling that can be any of the examples considered previously: a substitution quasiperiodic chain, the Penrose tiling or the Ammann-Beenker tiling. 
We are going to prove that this state has a non-trivial multifractal spectrum, and is therefore critical.

We consider the sequence of regions constructed by repetitively applying the tiling inflation rule $\sigma$ on a initial region $\reg_0$: $\reg_t = \sigma^t \reg_0$.

After a few algebraic manipulations of the above definitions, we arrive at
\begin{equation}
	d_q(\psi) = \frac{1}{q-1} \lim_{t \to \infty} \log \left( \frac{Z^{(t)}(2 \kappa)^q}{Z^{(t)}(2q\kappa)} \right) / \log Z^{(t)}(0)
\end{equation}
where $Z$ is the partition function already introduced in the specific case of the Fibonacci chain \eqref{eq:partition}.
Here we have used the important property that the height field is uncorrelated with the local configuration of the atoms. 

Furthermore, in all the cases considered here, the partition function has the scaling
\begin{equation}
	Z^{(t)}(\beta) \sim \omega^t(\beta).
\end{equation}
Therefore we arrive at the -- almost -- explicit expression of the fractal dimensions
\begin{equation}
	d_q(\psi) = \frac{1}{q-1} \log\left( \frac{\om(2 \kappa)^q}{\om(2 \kappa q)} \right) / \log(\om(0)).
\end{equation}
For the expression to be fully explicit, we need to compute $\om$ for the specific tiling at hand, as we already did in the Fibonacci case \eqref{eq:OmFibo}.

We can also compute the $f(\alpha)$ spectrum. 
Letting $x = 2\kappa$, we have
\begin{equation}
	\alpha_q = \log(\omega(x)) - x \frac{\omega'(qx)}{\omega(qx)}
\end{equation}
and
\begin{equation}
	f(\alpha_q) = \log(\omega(qx)) - qx \frac{\omega'(qx)}{\omega(qx)}
\end{equation}
Remark that whenever the function $\om$ is such that 
\begin{equation}
\label{eq:SymCond}
	\om(x) = e^{A x} \om(-x),
\end{equation}
the function $f(\alpha)$ is symmetric around its maximum.
Moreover, one can easily prove that the above condition \eqref{eq:SymCond} is equivalent to having a height distribution which is asymptotically symmetric around its maximum.
One can moreover express the maximum $h_M$ in terms of $A$: $h_M(t) = A t/2$


\section{Height distribution for 2D tilings}
\label{sec:2Dheights}

In this appendix, we compute the distribution of heights $N^{(t)}(h)$ on a region $\reg_t = \sub^t \reg_0$ inflated $t$ times, when $t \to \infty$.
For Penrose tiling, this distribution was first calculated by Sutherland in the saddle-point approximation in \cite{Sutherland}, while the exact results were later obtained in \cite{Repetowicz}.
We present here the detailed computation for the case of \AB\ tiling.

\begin{figure}[htp]
\centering
\includegraphics[width=0.4\textwidth]{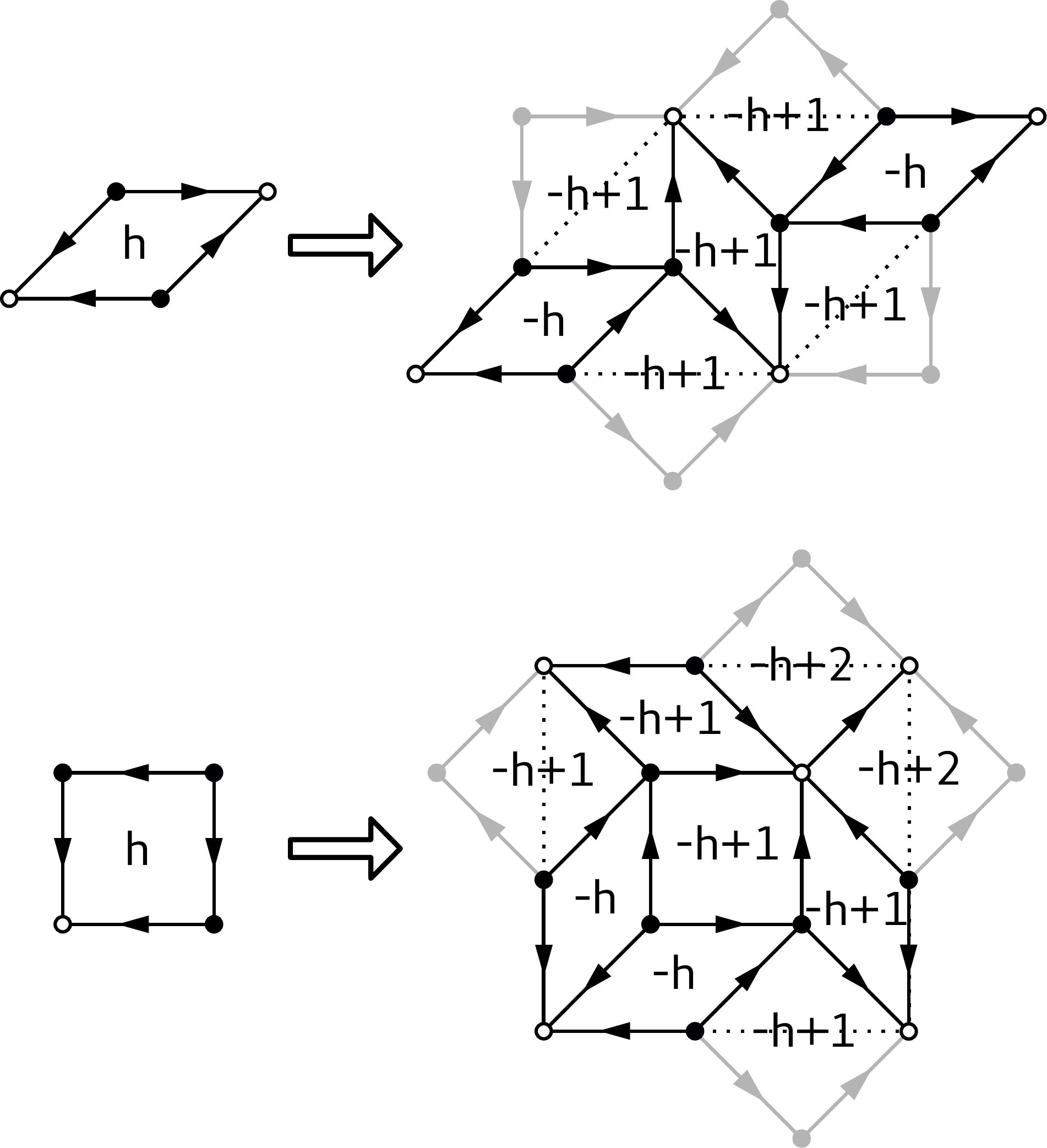}
\caption{The two tiles of the \AB\ tiling, and the supertiles obtained upon substitution. Reference vertices (see text for details) are marked by open circles. The height associated to each tile, i.e. the height on its reference site, is marked inside the tile.}
\label{fig:inflation}
\end{figure}

The \AB\ tiling can be built using a substitution rule (see fig.\eqref{fig:inflation}) which acts on the two tiles (lozenge and square).
Let $\mathbf{v} = (N_L, N_S)$ be a vector whose entries are respectively the number of lozenge and square tiles in a given region $\reg_0$.
Then the number of tiles in the inflated region $\reg_1 = \sub \reg_0$ is given by $M \mathbf{v}$ where $M$ is the \emph{inflation matrix}
\begin{equation}
\label{eq:inflation}
	M =
    \begin{bmatrix}
		3 & 4 \\
		2 & 3
	\end{bmatrix}.
\end{equation}

As in 1D, we now make use of the substitution to compute the distribution of heights.
Following Sutherland \cite{Sutherland}, we associate to each tile the height on one of its four vertices, taken as reference.
Here we choose the reference vertex to be the one which has arrows pointing in (see fig.~\eqref{fig:inflation}).
We now compute $N_{\mu}^{(t)}(h)$, the number of times height $h$ is reached on the reference site of the tiles of type $\mu = L~,S$.
As in 1D, the substitution reverses the effective arrows on the supertiles (see fig.~\eqref{fig:inflation}), meaning that height changes sign on the reference sites of the supertiles after a substitution.
As will be checked a posteriori, the heights are distributed in an \emph{environment-independent fashion}, meaning the arbitrariness in the choice of reference sites does not affect the result of the computation.
Introducing the vector $\mathbf{N}^{(t)}(h) = (N_L^{(t)}(h), N_S^{(t)}(h))$, one has
\begin{equation}
\label{eq:FPb}
	\mathbf{N}^{(t+1)}(-h) = \sum_{h'= 0}^2 M(h') \mathbf{N}^{(t)}(h-h'),
\end{equation}
where the \emph{generalized inflation matrices} are given by
\begin{equation}
	M(0) =
    \begin{bmatrix}
		2 & 2 \\
		0 & 0
	\end{bmatrix},
\end{equation}
\begin{equation}
	M(1) =
    \begin{bmatrix}
		1 & 2 \\
		2 & 2
	\end{bmatrix},
\end{equation}
\begin{equation}
	M(2) =
    \begin{bmatrix}
		0 & 0 \\
		0 & 1
	\end{bmatrix}.
\end{equation}
Note that the sum of the generalized inflation matrices, $M = \sum M(h')$, is just the inflation matrix $M$ of eq.~\eqref{eq:inflation}.
As 1D, the equation \eqref{eq:FPb} is a Fokker-Planck-like equation, and we expect the environment-specific height distributions to converge to Gaussian distributions in the large time limit.
Introducing partition function $\gen_\mu^{(t)}(\beta) = \sum_h N_\mu^{(t)}(h) \exp(\beta h)$, the evolution equation \eqref{eq:FPb} is recast to 
\begin{equation}
	\mathbf{\gen}^{(t+2)}(\beta) = \tilde{M}(-\beta)\tilde{M}(\beta) \mathbf{\gen}^{(t)}(\beta)
\end{equation}
where $\tilde{M}(\beta) = \sum_{h} M(h) \exp(-\beta h)$.
From this recursion relation, and since all the coefficients of $\tilde{M}$ are strictly positive, the Perron–Frobenius theorem applies, we deduce that the large time behavior of the partition function must be of the form
\begin{equation}
\label{eq:PartitionFunctionScaling}
	\gen_\mu^{(2t)}(\beta) \simop{t \to \infty} \omega^t(\beta) f_\mu(\beta)
\end{equation}
where $\om(\beta)$ is the largest eigenvalue of $\tilde{M}(-\beta) \tilde{M}(\beta)$, and $f_\mu(\beta)$ is the associated eigenvector.  
Explicit calculation gives
\begin{equation}
\label{eq:om}
    \om(\beta) = \frac{a(\beta) + \sqrt{a(\beta)^2 - e^{2 \beta}}}{e^\beta}
\end{equation}
with $a(\beta) = 4 \exp(2 \beta) + 9 \exp(\beta) + 4$.
Thus, in the $t \to \infty$ limit, the distributions all converge to the Gaussian distribution:
\begin{equation}
	P^{(t)}_\mu(h) \sim \frac{f_\mu}{\sqrt{4 \pi D t}} \exp\left(-\frac{h^2}{4 D t}\right)
\end{equation}
where $D$ is given by
\begin{equation}
	D = \frac{1}{6\sqrt{2}}.
\end{equation}

To obtain the fractal dimensions of the Penrose \SKK\ eigenstate we need a formula for $\omega(\beta)$, which was not explicitly calculated in \cite{Sutherland,Repetowicz}.
We give this formula without derivation since the computation is straightforward and most of it was already done in \cite{Repetowicz}:
\begin{equation}
	\om(\beta) = \frac{b(\beta) + \sqrt{b(\beta)^2 - 4e^{2\beta}}}{2}
\end{equation}
with $b(\beta) = \exp(2\beta) + 5\exp(\beta) + 1$.


\section{Variational method}
\label{app:var}

In this appendix we derive the variational equations used in Sec.~\ref{sec:var} for approximation of the 2D groundstate.

We recall that the Hamiltonian writes 
\begin{equation}
	H(t,V) = -t H_0 + V H_1,
\end{equation}
with 
\begin{equation}
	H_0 = \sum_{\langle i, j \rangle} c_j^\dagger c_i + \hc,
\end{equation}
and
\begin{equation}
	H_1 = \sum_i z_i c_i^\dagger c_i.
\end{equation}

We work with the following variational wavefunction:
\begin{equation}
	\psi(i) = C_{\mu(i)} e^{\kappa h(i)}
\end{equation}
where $\mu(i)$ is the nearest neighbors configuration of the site $i$: $\mu = $ A, B, C, D$_1$, D$_2$, E, F (see Sec.~\ref{sec:prefactor} for details).

\subsection{The variational energy}

We recall that variational method consists in minimizing the energy $E(\{C\},\kappa) = \Braket{\psi |H | \psi}/\Braket{\psi | \psi}$ with respect to variational parameters of $\ket{\psi}$, namely here the constant $\kappa$ and the 7 preexponential factors $C_\mu$, $\mu = $ A, B, C, D$_1$, D$_2$, E, F.
We thus have 8 variational equations, the solution of which gives an approximation to the exact groundstate wavefunction of our Hamiltonian.

\vspace{10pt}
\textbf{Evaluation of the norm $\Braket{\psi | \psi}$.}

In the following, we work on a finite-size sample $\reg_t$, and we then let the size of the sample go to infinity ($\reg_t \to \reg_\infty$).
In the same lines as in Sec.~\ref{sec:1Dheight} and in appendix~\ref{sec:2Dheights}, we first compute $N_\mu^{(t)}(h)$, the number of times we have height $h$ on sites of type $\mu$ inside region $\reg_t$.
For that, we introduce the generalized inflation matrix
\begin{equation}
	M(\beta) = \begin{bmatrix}
		1 & 1 & 1 & 1 & 0 & 0 & 0\\
		0 & 0 & 0 & 0 & 1 & 0 & 0\\
		0 & 0 & 0 & 0 & 0 & 1 & 0\\
		0 & 0 & 0 & 0 & 0 & 0 & 1\\
        0 & 0 & 0 & 0 & 0 & 0 & e^{\beta}\\
        0 & 0 & 0 & 0 & 2e^{\beta} & 3e^{\beta} & 2e^{\beta}\\
        8e^{\beta} & 8e^{\beta} & 8e^{\beta} & 8e^{\beta} & 5e^{\beta} & 2e^{\beta} & 0
	\end{bmatrix}.
\end{equation}
This matrix has the same physical interpretation as the generalized inflation matrices introduced in Sec.~\ref{sec:1Dheight} and in appendix.~\ref{sec:2Dheights}.
In particular, $M(0)$ is the geometrical inflation matrix relating the number of sites of type $\mu =$ A, B, \dots after $t$ inflations to the number of sites after $t+1$ inflations.
Let $f(\beta)$ is the eigenvector associated to the largest eigenvalue of $M(-\beta)M(\beta)$.
We have
\begin{align}
\Braket{\psi |\psi} &= \sum_{\mu} C_\mu^2 \sum_h e^{2\kappa h} N_\mu^{(t)}(h) \\
                           &= \sum_{\mu}  C_\mu^2  \gen_\mu^{(t)}( 2\kappa) \\
                           & = \omega^t(2 \kappa) \sum_\mu C_\mu^2 f_\mu(2\kappa)
\end{align}

\vspace{10pt}
\textbf{Evaluation of the average $\Braket{\psi|H_0|\psi}$.}

\begin{equation}
	\Braket{\psi |H_0 | \psi} = \sum_{\mu,\nu} C_\mu C_\nu \sum_h e^{\kappa( 2 h+\epsilon(\mu \to \nu))} N_\nu(\mu,h)
\end{equation}

$N_\nu(\mu,h)$ is the number of bonds $(\mu,\nu)$ with $\mu$ having height $h$.
$\epsilon(\mu \to \nu) = \pm 1$ respectively if the arrow goes from $\mu$ to $\nu$ or the reverse.
We can write $N_\nu(\mu,h)=z(\nu|\mu,h)N_\mu(h)$ with $z(\nu|\mu,h)$ the average number of type $\nu$ sites around type $\mu$ sites that have height $h$.
If the number of $\nu$ sites around $\mu$ is always the same, then $z(\nu|\mu,h)=z(\nu|\mu)$.
This is the case for $\mu < \nu$ (here we use lexicographic order: $1=A, 2=B, 3=C, 4=D_1, 5=D_2, 6=E, 7=F$).
Assuming $\mu < \nu$, we have
\begin{align}
	\sum_h e^{2\kappa h} N_\nu(\mu,h) & = z(\nu|\mu)\sum_h e^{2\kappa h} N_\mu(h) \\
	& = z(\nu|\mu)\gen_\mu(2\kappa)
\end{align}
Because the Hamiltonian is real symmetric, $e^{\kappa \epsilon(\mu \to \nu)} N_\nu(\mu,m)$ is symmetric under the exchange of $\mu$ and $\nu$. 
So, finally
\begin{equation}
	\Braket{\psi| H_0|\psi} = \om^t(2 \kappa) \sum_{\mu,\nu} C_\mu h_{\mu,\nu}(2\kappa) C_\nu
\end{equation}
where $h$ is the symmetric $7 \times 7$ matrix 
\begin{align}
	h_{\mu,\nu}(2 \kappa) & = e^{\kappa \epsilon(\mu \to \nu)} z(\nu|\mu) f_\mu(2 \kappa)~\text{if $\mu < \nu$} \\
               & = e^{\kappa \epsilon(\nu \to \mu)} z(\mu|\nu) f_\nu(2 \kappa)~\text{if $\mu > \nu$.}
\end{align}

\vspace{10pt}
\textbf{Evaluation of $\Braket{\psi|H|\psi}$}

This straightforwardly amounts to replacing $h_{\mu, \nu}$ by
\begin{equation}
	-t h_{\mu,\nu}(2\kappa) + V z_\mu f_\mu(2\kappa) \delta_{\mu,\nu}
\end{equation}
with $z_\mu$ the coordination of type $\mu$ sites.

\subsection{The variational equations}
If the energy has an extrema with respect to $p$, it obeys the equation
\begin{equation}
	 \partial_p \Braket{\psi|H|\psi} = E(p) \partial_p \Braket{\psi|\psi}.
\end{equation}
Here we have two kinds of parameters: the preexponential factors $C$ and $\kappa$. Let us consider each in turn.

\vspace{10pt}
\textbf{Extremization with respect to $C$}

Let us first extremize for $H_0$. Since $h$ is symmetric, we have
\begin{equation}
\label{eq:min_c}
	\sum_\nu h_{\mu,\nu} C_\nu = E f_\mu C_\mu
\end{equation}
So, $\mathbf{C}(\kappa)$ is an eigenvector of the matrix $M_{\mu, \nu} = h_{\mu,\nu}(\kappa)/f_\mu(\kappa^2)$, with eigenvalue $E$.
There are thus 7 independent solutions for $\mathbf{C}(\kappa)$, for each value of $\kappa$. 
The extension to $H$ is simple: $M_{\mu, \nu}$ becomes
\begin{equation}
	M_{\mu,\nu} = -t h_{\mu,\nu}(2\kappa)/f_\mu(2\kappa) + V z_\mu \delta_{\mu,\nu}
\end{equation}
Finding the extrema with respect to the $C$ parameters amounts to diagonalizing the matrix $M$. 
Although it is possible to diagonalize $M$ exactly, we do not reproduce the solution here as it is too long.

\vspace{10pt}
\textbf{Extremization with respect to $\kappa$}

Since $h$ is symmetric, we can write 
\begin{equation}
	\Braket{\psi|H_0|\psi} = 2 \sum_{\mu < \nu} C_\nu z(\nu|\mu)f_\mu(2\kappa)e^{\kappa \epsilon(\mu \to \nu)} C_\mu
\end{equation}
Then, extremization yields
\begin{equation}
\label{eq:min_beta}
	\sum_{\mu < \nu} C_\nu z(\nu|\mu)\partial_\kappa \left (f_\mu(2\kappa)e^{\kappa \epsilon(\mu \to \nu)} \right ) C_\mu = E \sum_\mu C_\mu^2 f'_\mu(2\kappa)
\end{equation}
The extension to $H$ is, again, straightforward.
We were not able to solve this last equation analytically. 
We instead solved it numerically.

\section{Is the groundstate of 1D chains an \SKK\ state?}

In this section, we consider the simple tight-binding model introduced in part 1:
\begin{equation}
\label{eq:ham}
	H =- \sum_{i} t_i c_{i+1}^\dagger c_i + \hc.
\end{equation}
We recall that the hopping amplitude $t_i$ takes the value $t_a$ or $t_b$ if the letter $i$ in the quasiperiodic chain we consider is an $a$ or a $b$.
In part 1 we considered in details the case of the Fibonacci chain \eqref{eq:FiboHam}, and we also considered more generally chains of the \emph{metallic mean sequence}. 
We showed that the $E=0$ state on these chains is an \SKK\ state. 
It is natural to wonder whether the \emph{groundstate} on these chains is also an \SKK\ state.
We show in this appendix that it is not the case, in the neighborhood of the periodic chain.

We introduce the \emph{discrete logarithmic derivative}
\begin{equation}
	\dlogpsi(i) = \frac{\psi(i) - \psi(i-1)}{\psi(i)},
\end{equation}
where $\psi$ is the groundstate wavefunction.
Assume that the groundstate is of the \SKK\ type, i.e. $\psi(i) = \psi_0(i) \exp(\kappa h(i))$.
Assume moreover that we are close to the periodic chain (i.e. $t_a \simeq t_b$).
Then $\kappa \ll 1$ and we have
\begin{equation}
	\dlogpsi(i) = \kappa A(i-1 \to i) + \dlogpsi_0(i),
\end{equation}
where $A(i-1 \to i) = h(i) - h(i-1)$ is the arrow function associated to the height field (see Part 1), and where $\dlogpsi_0$ is the discrete logarithmic derivative of the preexponential factor $\psi_0$.
Now, the arrow function is \emph{local} (i.e. the arrow only depends on the local arrangement of the atoms), and the preexponential factor is local as well, by definition. 
Thus, if the groundstate is of the \SKK\ type, its logarithmic derivative $\dlogpsi$ must be local. 
That is to say, \emph{in perpendicular space, $\dlogpsi$ must exhibit plateaus}, just like the preexponential factor of the groundstate for 2D tilings, see eg. \eqref{fig:preexp}.
$\dlogpsi$ is thus a tool to determine whether the groundstate is of the \SKK\ type or not.

\begin{figure}[htp]
	\centering
	\includegraphics[width=0.4\textwidth]{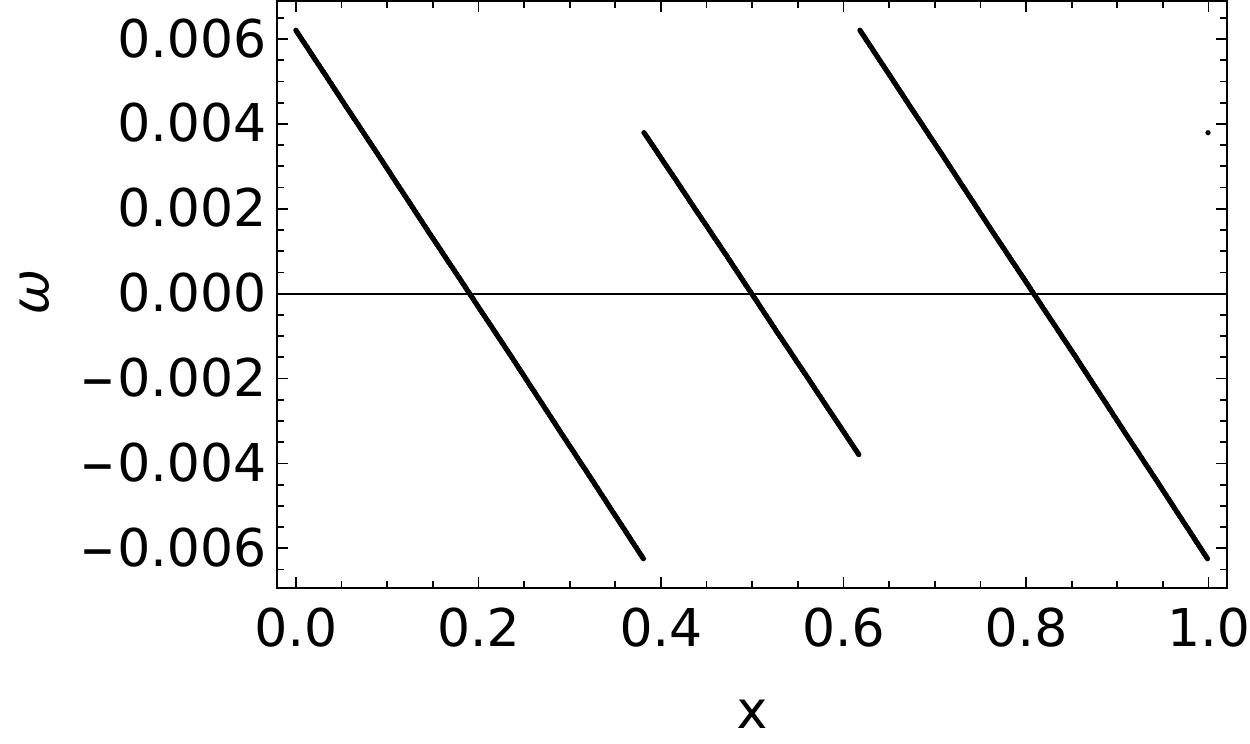}
	\caption{Discrete logarithmic derivative of the groundstate of the Fibonacci chain, as a function of the perpendicular space coordinate. Here $t_b/t_a=0.99$. Numerical computation was performed on a chain of 1597 sites.}
	\label{fig:logderivative}
\end{figure}
Figure \eqref{fig:logderivative} shows the logarithmic derivative computed numerically, for a chain close to the periodic chain ($t_b/t_a=0.99 \simeq 1$).
Instead of the plateaus expected if the ground state were of the \SKK\ type, the logarithmic derivative is piecewise linear.
We thus conclude that the ground state cannot be of the \SKK\ type.
If we move further away from the periodic chain, the lines observed on fig.~\eqref{fig:logderivative} become irregular, showing a devil's staircase structure.

\bibliography{bib.bib}{}
\bibliographystyle{plain}

\end{document}